\documentclass[aps,prb,preprint,showpacs]{revtex4}
\usepackage{bm}
\usepackage{epsfig}
\usepackage{color}
\usepackage{graphicx}
\usepackage{amssymb}
\usepackage{amsmath}
\usepackage{wrapfig}
\usepackage[center]{caption}
\usepackage[utf8]{inputenc}

\begin{document}

\title{Spin-orbit interactions may relax the rigid conditions leading to flat 
bands}

\author{N\'ora~Kucska and Zsolt~Gul\'acsi}
\affiliation{Department of Theoretical Physics, University of
Debrecen, H-4010 Debrecen, Bem ter 18/B, Hungary}
\date{\today }
     
\begin{abstract} 
Flat bands are of extreme interest in a broad spectrum of fields since given 
by their high degeneracy, a small perturbation introduced in the system is able
to push the ground state in the direction of an ordered phase of interest. 
Hence the flat band engineering in real materials attracts huge attention.
However, manufacturing a flat band represents a difficult task
because its appearance in a real system is connected to rigid mathematical 
conditions relating a part of Hamiltonian parameters. Consequently,
whenever a flat band is desired to be manufactured, these Hamiltonian 
parameters must be tuned exactly to the values fixed by these rigid 
mathematical conditions. Here we demonstrate that taking the 
many-body spin-orbit interaction (SOI) into account-- which can be continuously
tuned e.g. by
external electric fields --, these rigid mathematical conditions can be
substantially relaxed. On this line we show that a $\sim 20-30 \%$ variation 
in the Hamiltonian parameters rigidly fixed by the flat band conditions can 
also lead to flat bands in the same or in a bit displaced position on the 
energy axis. This percentage can even increase to $\sim 80 \%$ in the presence 
of an external magnetic field. The study is made in the case of conducting
polymers. These systems are relevant not only
because they have broad application possibilities,
but also because they can be used to present the mathematical background
of the flat band conditions in full generality, in a concise, clear and 
understandable manner applicable everywhere in itinerant systems.   
\end{abstract}


\maketitle

\section{Introduction}

Flat bands are attracting great interest today given by the broad application
possibilities of the huge degeneracy they provide. Indeed, flat bands
appear in several circumstances as:
chiral edge mods broadband topological slow light \cite{E1}, 
diffraction-free photonics via collective 
excited states of atoms in Creutz super-radiance lattices \cite{E2},
interaction-enhanced group velocity in optical Kagome lattices \cite{E3},
production of topological states in 1D optical lattices 
\cite{E4}, generation of flat bands in non-Hermitian optical lattices
\cite{E5}, realization of tilted Dirac cones from flat-bands which lead to 
intricate transport phenomena \cite{E6}, engineering flat band PT symmetric
meta-materials \cite{E7}, use of singularities emerging on flat bands \cite{E8},
artificial flat band systems \cite{E9}, SQUID meta-materials on Lieb lattices
\cite{E10}, or are of interest because of the emergence of different ordered 
phases in flat band systems as
superconductivity \cite{E11}, ferromagnetism \cite{E12}, semimetal magnetic
ordering \cite{E13}, excitonic insulator \cite{E14}, etc. Flat bands also 
produce interesting effects as quantized circular photo-galvanic effect 
\cite{E15}, ordered quantum dot arrays formed by moire excitons \cite{E16},
emergence of non-contractible-loop-states \cite{E17}, etc.

Flat bands occur in several type of materials from which conducting polymers
 \cite{E13,E18,E19,E20,E21,E22}
have broad application possibilities covering
thermal conductivity enhancement \cite{E23}, carrier charge transport 
\cite{E24}, heat exchangers and energy storage \cite{E25}, soft high 
performance capacitors \cite{E26}, switchers and commutators \cite{E27},
sensors \cite{E28}, high performance batteries \cite{E29}, biodegradable
plastics \cite{E30}, light-emitting diodes \cite{E31}, organic transistors
\cite{E32}, and even life sciences and medicine \cite{E33,E34,E35}. This is 
the reason why in the study of flat band characteristics, we
exemplify the observed properties in the case of conducting polymers. 

Flat bands can be effective \cite{E20,E36,E37}, or bare (i.e. provided
exclusively by $\hat H_{kin}$, kinetic energy part of the Hamiltonian).
Their main source of difficulties is that they are determined by rigid 
mathematical conditions
connected to the parameters (e.g. hopping matrix elements, coupling constants) 
of the Hamiltonian ($\hat H$). Indeed, deducing the band structure 
in a lattice, in
principle we obtain from the one particle part of the Hamiltonian a secular 
equation of the form
\begin{eqnarray}
Q(\epsilon, \{p_i\}, \{\text{trig}_j({\bf k}{\bf x}_{\alpha})\})=0
\label{equ1}
\end{eqnarray}
where $\epsilon=E_n({\bf k})$ provides the energy spectrum, $ \{p_i\}$ 
represents
the set of the parameters of the Hamiltonian ($i=1,2,...m_{max}$), 
${\bf x}_{\alpha}$ are the Bravais 
vectors of the lattice, $\text{trig}_j(z)$ represent trigonometric functions of
$\sin(nz)$, $\cos(nz)$ type (where $n$ is an integer) holding in their argument 
the ${\bf k}$ momentum dependence. The notation  
$\{\text{trig}_j({\bf k}{\bf x}_{\alpha})\}$ 
represents the set of all trigonometric functions emerging in the secular 
equation Eq.(\ref{equ1}). In  Eq.(\ref{equ1}) all trigonometric contributions
$\text{trig}_j$ emerge in Q additively, with multiplicative coefficients 
$T_j(\{p_i\})$ 
[i.e. as  $V_j=T_j(\{p_i\}) \text{trig}_j({\bf k}{\bf x}_{\alpha})$] which depend 
on the Hamiltonian parameters $\{p_i\}$. Eliminating these coefficients
\begin{eqnarray}
T_j(\{p_i\})=0, \quad j=1,2,3,...,m
\label{equ2}
\end{eqnarray}
the ${\bf k}$ dependence disappears from the secular equation Eq.(\ref{equ1}),
hence from $Q =0$ we find ${\bf k}$ independent $\epsilon$ values, i.e. flat
bands [see for exemplification Eq.(\ref{equ13}-\ref{equ15})].
As seen from Eq.(\ref{equ2}), when flat bands emerge, interdependences 
between Hamiltonian parameters must be present. If in  Eq.(\ref{equ2}) one has
$j=1,2,...m < m_{max}$, these interdependencies rigidly fix the value of $m$ 
Hamiltonian parameters. Hence, when a flat band appears, only $(m_{max}-m)$
Hamiltonian parameters can be arbitrarily chosen, and $m$ Hamiltonian parameters
remain rigidly fixed (given and determined by the arbitrarily taken 
$(m_{max}-m)$ $\hat H$ independent parameters). One mentions that Q in
Eq.(\ref{equ1}) contains additively also a $V_0=T_{j=0}$ term which does not 
contain ${\bf k}$, and explicitly one has $Q=\sum_{j=0}^m V_j$, [see also Eq.
(\ref{equ25})]. 

Furthermore if the $T_j$ coefficients in Q also contain the parameter
$\epsilon$, we can fix the origin of the energy axis to the position 
of the flat band
(i.e. $\epsilon=0$), and the deduction of the flat band conditions can be 
similarly treated, as presented above in Eqs.(\ref{equ1},\ref{equ2}).
 
Usually, the flat band conditions Eq.(\ref{equ2}) are deduced from a given 
$\hat H$ describing itinerant systems with independent orbital and spin degrees
of freedom. This state of facts is motivated by the observation that the 
many-body
spin-orbit interaction $\hat H_{SO}=\lambda {\bf \sigma} \cdot (\nabla V \times 
{\bf k})$ is usually small. Here ${\bf \sigma}$ represents the spin of carriers,
${\bf k}$ is their momentum, $\nabla V$ the potential gradient, while 
$\lambda << 1$ is the strength of the spin-orbit interaction.
When the system is interacting, (e.g. the leading term of the Coulomb 
interaction in a many-body 
system, the on-site Coulomb repulsion $U > 0$ is present), the use of 
$H_{SO}$ introduces 
supplementary complications since because $\lambda << U$ even the perturbative 
treatment is questionable, hence enforcing special treatment for 
obtaining exact results \cite{E38,E39,E40}.

Even if the spin-orbit 
interaction (SOI) is small, its effect is major since it breaks the 
spin-projection double degeneracy of each band \cite{E41}, and leads to 
several interesting
effects as: stable soliton complexes \cite{E42}, enhances transport properties
\cite{E43}, influences graphene properties \cite{E44}, coupling of Hofstadter 
butterfly pairs \cite{E45}, topological excitations \cite{E46}, provides
stripe and plane-wave phases \cite{E47}, is able to produce spin-memory loss
\cite{E48}, influences proximity effects at interfaces \cite{E49}, leads to
anomalous Josephson effect \cite{E50} and condensed phases \cite{E38}.
Furthermore, in several circumstances $\lambda$ is strongly tunable \cite{E51},
can be enhanced by Coulomb correlations \cite{E52}, can be increased by
doping \cite{E53}, 
structural conformation (e.g. altering torsion in conjugated polymers) 
\cite{E54a}, twist of the aromatic rings along the conjugation path \cite{E54b},
and can be even tuned by external electric field \cite{E54}.

In this paper we show that taking into account $\hat H_{SO}$ in the system
Hamiltonian $\hat H$, the rigid flat band conditions in Eq.(\ref{equ2}) can be 
substantially relaxed. This procedure is tempting because the strength of SOI
can be continuously tuned by an applied external electric field.
Consequently, engineering a flat band in a real system
is in fact more easily achievable compared to how it was 
considered before.
As we mentioned previously, we exemplify our results on conducting polymers. 
Two spin-orbit couplings are considered, one (denoted by $\lambda$) as in base,
and another one (denoted by $\lambda_c$) as inter-base contribution. In order to
obtain more information, also external magnetic field is considered acting via
Peierls phase factors. For the conducting polymer a pentagon chain is 
considered (e.g. polyaminotriazole type of chain) since this was one of the 
first produced conducting polymer. 

The remaining part of the paper is constructed as follows: Sect. II. presents 
the studied system, Sect. III. deduces the band structure, and determines the
flat bands, Sect. IV. (Sect. V.) describes how the mathematically rigid flat 
band conditions can be relaxed by spin orbit interactions maintaining (not
maintaining) the position of the flat band, Sect. VI. summarizes the paper, and
finally, Appendices A,B,C,D,E, containing mathematical details, close the 
presentation.

\section{The system studied}

A schematic plot of the
unit cell of the system containing 6 sites is presented in Fig.1. 
The upper antenna in the pentagon chain (as e.g. in polyaminotriazole, see 
Fig.2) is considered simply as the bonds (5,6) on Fig.1, since this structure 
is able to describe qualitatively correct its effect in the band structure.
The external magnetic field is perpendicular to the plane of the cell. At the 
level of the Hamiltonian the system is described by 
\begin{eqnarray}
\hat H = \hat H_{kin}+ \hat H_{SO}
\label{equ3}
\end{eqnarray}
where, denoting by $n=1,2,...,6$ the in-cell position of atoms, $\hat H_{kin}$ 
is given by
\begin{figure}
	\includegraphics[width=7cm]{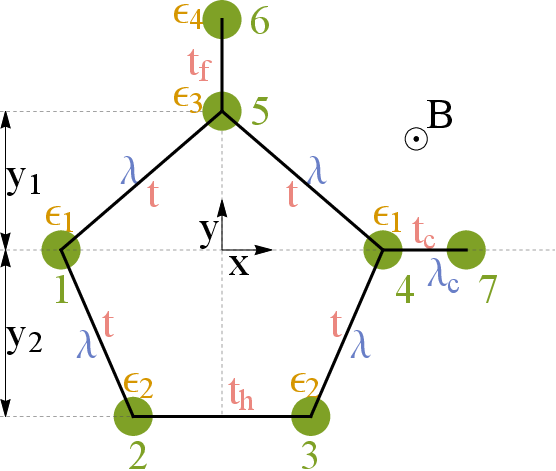}
	\caption{The pentagonal unit cell, with the nearest 
neighbour hopping matrix elements ($t, t_h, t_c, t_f$), the Rashba 
couplings ($\lambda, \lambda_c$), the on-site one-particle potentials 
($\epsilon_1,\epsilon_2, \epsilon_3, \epsilon_4$) and the external 
magnetic field ($B$). }
\label{fig1}
\end{figure}
\begin{figure}
	\includegraphics[width=5cm]{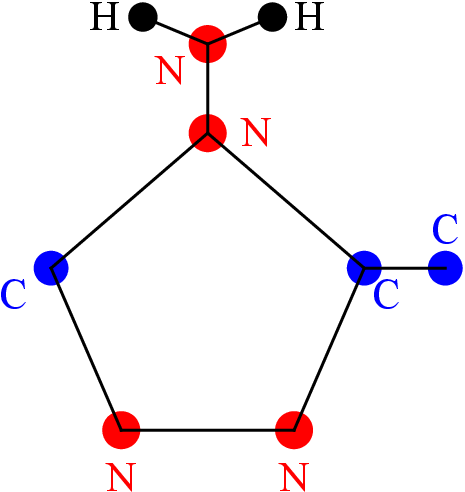}
	\caption{Schematic plot of the polyaminotriazole cell}
	\label{fig2}
\end{figure}

\begin{eqnarray}
\hat H_{kin}&=& \sum_{i,\sigma} [( t e^{i\phi_{1,5}}\hat c^{\dagger}_{i,1,\sigma}
\hat c_{i,5,\sigma'} + t e^{i\phi_{2,1}} \hat c^{\dagger}_{i,2,\sigma} \hat c_{i,1,\sigma'} 
+ t e^{i\phi_{4,3}} \hat c^{\dagger}_{i,4,\sigma} \hat c_{i,3,\sigma'} + 
t e^{i\phi_{5,4}}\hat c^{\dagger}_{i,5,\sigma} \hat c_{i,4,\sigma'} 
\nonumber\\ 
&+& t_c e^{i \phi_{7,4}} \hat c^{\dagger}_{i+a,7,\sigma} \hat c_{i,4,\sigma'} 
+ t_h e^{i \phi_{3,2}} \hat c^{\dagger}_{i,3,\sigma} \hat c_{i,2,\sigma} +  
t_f e^{i \phi_{6,5}}\hat c^{\dagger}_{i,6,\sigma} \hat c_{i,5,\sigma} + H.c. ) 
\nonumber\\
&+&
\sum_{n=1}^6\epsilon_{n} \hat c^{\dagger}_{i,n,\sigma} \hat c_{i,n,\sigma}].
\label{equ4}
\end{eqnarray}
Here $c^{\dagger}_{i,n,\sigma}$ creates an electron with $\sigma$ spin projection
in the $n$ position of the cell placed at the site $i$; the $t, t_h, t_c, t_f$ 
are nearest neighbour hopping matrix elements; while $\epsilon_n$ are 
the on-site
one-particle potentials at the in-cell positions $n$. Based on the symmetry of 
the unit cell, one uses the notations $\epsilon_1=\epsilon_{n=1}=\epsilon_{n=4},
\epsilon_2=\epsilon_{n=2}=\epsilon_{n=3}, \epsilon_3=\epsilon_{n=5}, \epsilon_4=
\epsilon_{n=6}$. The Peierls phase factors
$\phi_{n,n'}$ (describing the effect of the external magnetic field on the 
orbital motion of the carriers) are deduced in the Appendix A. Based on the
obtained results, one uses the following notations
$\phi_{3,2}=\phi_1, \phi_{4,3}=\phi_{2,1}=\phi_2, \phi_{5,4}=\phi_{1,5}=\phi_3,
\phi_{5,6}=\phi_{7,4}=0$.

Concerning $H_{SO}= \lambda {\bf \sigma} \cdot (\nabla V \times {\bf k})$, it 
introduces spin-flip type hoppings along the bonds of the system \cite{E38}.
Since spin-orbit coupling for carbon influences considerably the physical
processes in carbon made materials \cite{E55,E56}, we 
take into consideration $H_{SO}$ on bonds containing carbon atoms. This choice
is supported also by the fact that these bonds provide the conjugated (i.e. 
conducting) nature of the polymer. From these bonds two manifolds can be 
constructed: in-cell bonds [(1,5);(2,1);(4,3);(5,4); see Fig.1], and inter-cell
bonds [(7,4) in Fig.1]. Since the strength of the spin-orbit coupling on
inter-cell bonds can be increased by atom intercalation \cite{E57} and the
ending atoms on these bonds are different from the ending atoms on in-cell
bonds, the SOI
coupling on these bonds will be denoted by $\lambda_c$, while the in-cell
SOI coupling by $\lambda$. In these conditions, taking into account Rashba
interaction in polymers \cite{E54}, $H_{SO}$ becomes
\begin{eqnarray}
\hat H_{SO} &=& \sum_{i,\sigma} ( t^{\sigma,-\sigma}_{1,5} \hat c^{\dagger}_{i,1,\sigma}
\hat c_{i,5,-\sigma} + t^{\sigma,-\sigma}_{2,1} \hat c^{\dagger}_{i,2,\sigma} \hat c_{i,1,
-\sigma} + t^{\sigma,-\sigma}_{4,3} \hat c^{\dagger}_{i,4,\sigma} \hat c_{i,3,-\sigma} + 
t^{\sigma,-\sigma}_{5,4} \hat c^{\dagger}_{i,5,\sigma} \hat c_{i,4,-\sigma} 
\nonumber\\
&+& 
t_c^{\sigma,-\sigma} \hat c^{\dagger}_{i+a,7,\sigma} \hat c_{i,4,-\sigma} +H.c.),
\label{equ5}
\end{eqnarray}
where $\lambda= t_{5,1}^{\uparrow,\downarrow} =  t_{1,5}^{\downarrow, \uparrow} = 
t_{1,2}^{\downarrow,\uparrow} = t_{2,1}^{\uparrow,\downarrow}
= t_{3,4}^{\downarrow, \uparrow} = t_{4,3}^{ \uparrow, \downarrow} = 
t_{4,5}^{\uparrow, \downarrow} = t_{5,4}^{\downarrow, \uparrow}$, and
$\lambda_c = t_c^{\uparrow,\downarrow}=t_{7,4}^{\uparrow,\downarrow}$, 
furthermore 
$t_{i,j}^{\uparrow,\downarrow}=-t_{j,i}^{\uparrow,\downarrow}$ holds.

As mentioned previously the strength of $H_{SO}$ can be continuously tuned by
an applied external electric field \cite{E58,E59}. One applies the external 
${\bf E}= E {\vec k}$ field in the $z$ direction (perpendicular to the plane of 
the chain, ${\vec k}$ being the unit vector in $z$ direction).
Since the carriers move in the $x$ direction (see Fig.3), the first quantized
Rashba Hamiltonian becomes $\hat H_R = - i \eta \sigma_y k_x$ \cite{E60,E54}, 
(here $k_x$ is the momentum along the $x$ axis, i.e. along the polymer chain),
hence
the spin is oriented along the $y$ axis. After this step if one couples the 
external magnetic field ${\bf B}$ along the $z$ axis, since the magnetic 
induction
and the spin vector are perpendicular, the Zeeman term provides zero 
contribution, and the external magnetic field acts only via the Peierls phase 
factor. If the source of SOI is exclusively the external electric field, 
$\lambda=\lambda_c$, and the connection of $\lambda$ to $E$ is given by
\cite{E61}
\begin{eqnarray}
\lambda = \bar K E, \quad \bar K = \frac{|q|\hbar^2}{4m^2c^2}\frac{2\pi}{
\lambda_D},
\label{equ6}
\end{eqnarray} 
where in the expression of the coefficient $\bar K$, $q$ and $m$ are the charge
and (rest) mass of the carriers, $\lambda_D$ is their de Broglie wavelength, 
and $c$ is the speed of light.

\section{The band structure}

\begin{figure}
	\includegraphics[width=7cm]{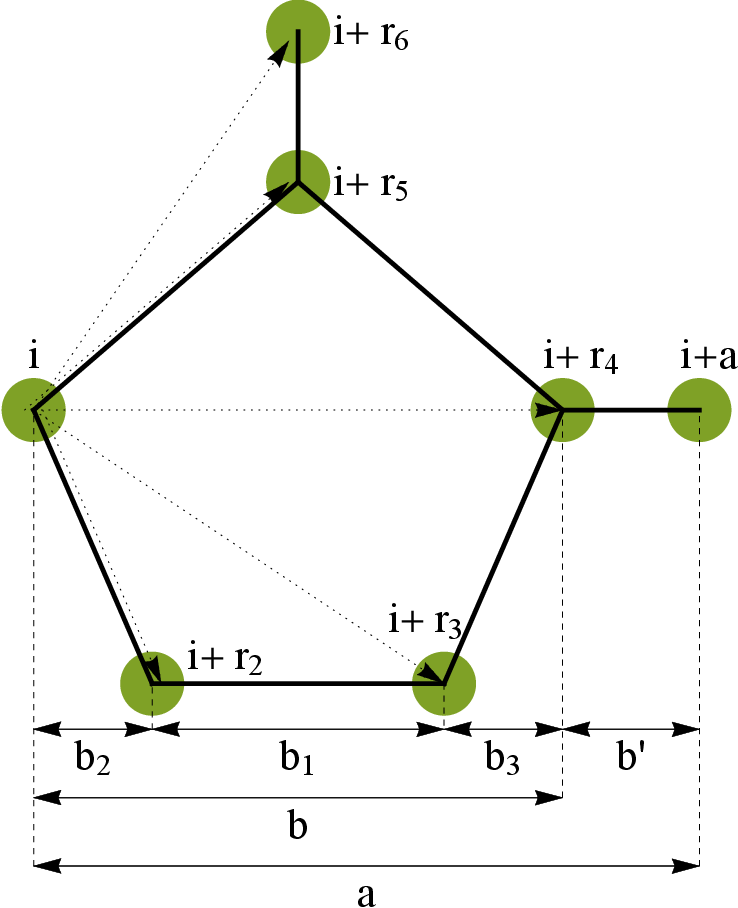}
	\caption{Notations used for the pentagonal unit cell}
	\label{fig2}
\end{figure}

First we transform  the $\hat H$ Hamiltonian from Eq.(\ref{equ3})
to $k$-space. The fermionic operators are Fourier transformed via
$\hat{c}_{i,r_n,\sigma}=\frac{1}{\sqrt{N_c}}\sum_k e^{-ik(i+r_n)}\hat{c}_{n,k,\sigma}$,
where $N_c$ represents the number of unit cells and 
$k$ is directed along the $x$ axis (see Fig.3). One obtains [see also Eq.(A4)]
\begin{eqnarray}
&&\hat H = \sum_k \sum_{\sigma,\sigma'} [t^{\sigma,\sigma'}_{1,5} \hat c^{\dagger}_{
\bf{k} ,1,\sigma}
\hat c_{\bf{k} ,5,\sigma'}  e^{i {\bf{k (r_1-r_5}})} + t^{\sigma,\sigma'}_{2,1} 
\hat c^{\dagger}_{\bf{k} ,2,\sigma} \hat c_{i,1,\sigma'}  e^{i {\bf{k (r_2-r_1}})} + 
t^{\sigma,\sigma'}_{4,3}
\hat c^{\dagger}_{\bf{k} ,4,\sigma} \hat c_{\bf{k} ,3,\sigma'}  e^{i {\bf{k (r_4-r_3}})} 
\nonumber\\
&& + t^{\sigma,\sigma'}_{5,4} \hat c^{\dagger}_{\bf{k} ,5,\sigma} \hat c_{\bf{k} ,4,\sigma'} 
e^{i {\bf{k (r_5-r_4}})}  +  t_c^{\sigma,\sigma'} \hat c^{\dagger}_{\bf{k} ,1,\sigma} 
\hat c_{\bf{k} ,4,\sigma'}  e^{i {\bf{k (a-r_4}})} +
t_h \hat c^{\dagger}_{\bf{k} ,3,\sigma} \hat c_{\bf{k} ,2,\sigma}  e^{i {\bf{k (r_3-r_2}})} 
\nonumber\\
&& + t_f \hat c^{\dagger}_{\bf{k} ,6,\sigma} \hat c_{\bf{k} ,5,\sigma}  
e^{i {\bf{k (r_6-r_5}})} + h.c.] +
\sum_{n}\epsilon_n \hat c^{\dagger}_{\bf{k} ,n} \hat c_{\bf{k} ,n}.
\label{equ7}
\end{eqnarray}
Here $r_n$ represents the in-cell position of the $n$-th atoms in the cell, 
and $r_1=0$ is considered. The terms in the exponents are obtained via 
(see Fig.3):
\begin{eqnarray}
&& {\bf{k (r_4-r_3}})={\bf{k (r_2-r_1}})  = k b_2, \quad {\bf{k (r_6-r_5}}) = 0,
\quad {\bf{k (a-r_4}}) = kb', 
\nonumber\\
&& {\bf{k (r_3-r_2}}) = kb_1, \quad
{\bf{k (r_5-r_4}}) = {\bf{k (r_1-r_5}}) = \frac{kb}{2}.
\label{equ8}
\end{eqnarray}
Using Eq.(\ref{equ8}) in Eq.(\ref{equ7}) one finds
\begin{eqnarray}
&&\hat H = \sum_k \sum_{\sigma,\sigma'} [t^{\sigma,\sigma'}_{1,5} \hat c^{\dagger}_{\bf{k},
1,\sigma}
\hat c_{\bf{k} ,5,\sigma'}  e^{i \frac{kb}{2}} + t^{\sigma,\sigma'}_{2,1} \hat c^{\dagger}_{
\bf{k} ,2,\sigma} \hat c_{i,1,\sigma'}  e^{i k b_2} + t^{\sigma,\sigma'}_{4,3}
\hat c^{\dagger}_{\bf{k} ,4,\sigma} \hat c_{\bf{k} ,3,\sigma'}  e^{i k b_2} 
\nonumber\\
&& + t^{\sigma,\sigma'}_{5,4} \hat c^{\dagger}_{\bf{k} ,5,\sigma} \hat c_{\bf{k} ,4,\sigma'} 
e^{i  \frac{kb}{2}} +
t_h \hat c^{\dagger}_{\bf{k} ,3,\sigma} \hat c_{\bf{k} ,2,\sigma}  e^{i kb_1}
+ t_f \hat c^{\dagger}_{\bf{k} ,6,\sigma} \hat c_{\bf{k} ,5,\sigma} 
\nonumber\\
&& + t_c^{\sigma,\sigma'} \hat c^{\dagger}_{\bf{k} ,1,\sigma} \hat c_{\bf{k} ,4,\sigma'}  
e^{i  kb'} + h.c.] +
\sum_{n}\epsilon_n \hat c^{\dagger}_{\bf{k} ,n} \hat c_{\bf{k} ,n} .
\label{equ9}
\end{eqnarray}
One observes that $\hat H$ in Eq.(\ref{equ9}) can be written as
\begin{eqnarray*}
\hat H = 
\sum_k [ (\hat{c}^{\dagger}_{k,1,\uparrow}, \ldots , \hat{c}^{\dagger}_{k,6,\uparrow},
\hat{c}^{\dagger}_{k,1,\downarrow}, \ldots, \hat c^{\dagger}_{k,6,\downarrow} )  \bold{M} 
\begin{pmatrix}
\hat c_{k,1,\uparrow} \\ \vdots  \\  \hat c_{k,6,\uparrow} \\ \hat c_{k,1,\downarrow} \\
\vdots \\ \hat c_{k,6,\downarrow}
\end{pmatrix}, 
\end{eqnarray*}
where $\bold{M}$, being a $12\times 12$ matrix, can be written in the 
following form:
\begin{equation}
\bold{M}= 
	\begin{pmatrix}
	\bold{M}_1 & \bold{M}_2\\
	\bold{M}_3 & \bold{M}_4
	\end{pmatrix} .
\label{equ10}
\end{equation}
Here, the $\bold{M}_j$, $j=1,2,3,4$ contributions are given as follows:
\setcounter{MaxMatrixCols}{12}

\begin{equation*}
\bold{M}_1 = 
	\begin{pmatrix}
	\epsilon_1 &  t e^{-i(kb_2 + \varphi_2)} & 0  & t_c  e^{ikb'} & t_{1,5}  
e^{-i(\frac{kb}{2}-\varphi_3)}  & 0 \\
	
	t e^{i(kb_2+ \varphi_2)} & \epsilon_2 & t_h e^{-i(kb_1+\varphi_1)}  & 0 & 0  & 0
\\
	
	0 & t_h e^{i(kb_1+\varphi_1)} & \epsilon_2  & t e^{-i(kb_2+\varphi_2)} & 0  & 0 
\\
	
	t_c e^{-ikb'} & 0 &  t e^{i(kb_2+\varphi_2)}   & \epsilon_1  & t e^{
i(\frac{kb}{2}-\varphi_3)}   & 0 \\
	
	t  e^{i(\frac{kb}{2}-\varphi_3)}  & 0 & 0  & t  e^{-i(\frac{kb}{2}-\varphi_3)} & 
\epsilon_3  & t_f \\
	
	0 & 0 & 0  & 0 & t_f  & \epsilon_4 
	\end{pmatrix}
        \end{equation*}

\begin{equation*}
\bold{M}_2=
	\begin{pmatrix}
 0  & -\lambda e^{-i(kb_2+\varphi_2)} & 0  & \lambda_c e^{ikb'} & -\lambda 
e^{-i(\frac{kb}{2}-\varphi_3)} & 0\\
	
 \lambda e^{i(kb_2+\varphi_2)}   & 0 & 0  & 0 & 0 & 0\\
	
 0  & 0 & 0  & -\lambda e^{-i(kb_2+\varphi_2)} & 0 & 0\\
	
 -\lambda_c  & 0 & \lambda  e^{i(kb_2+\varphi_2)}   & 0 & \lambda 
e^{i(\frac{kb}{2}-\varphi_3)}  & 0\\
	
\lambda  e^{-(\frac{kb}{2}-\varphi_3)}   & 0 & 0  & -\lambda e^{-i(\frac{kb}{2}-\varphi_3)}
& 0 & 0\\
	
 0  & 0 & 0  & 0 & 0 & 0
\end{pmatrix}
\end{equation*}

\begin{equation*}
\bold{M}_3=
	\begin{pmatrix}	
	0 & \lambda e^{-i(kb_2+\varphi_2)} & 0  & -\lambda_c  e^{ikb'}  & \lambda  
e^{-i(\frac{kb}{2}-\varphi_3)}  & 0\\
	
	-\lambda e^{i(kb_2 + \varphi_2)} & 0 & 0  & 0 & 0  & 0\\
	
	0 & 0 & 0  & \lambda e^{-i(kb_2+\varphi_2)}  & 0  & 0 \\
	
	\lambda_c e^{-ikb'} & 0 & -\lambda  e^{i(kb_2 + \varphi_2)}   & 0 & -\lambda 
e^{i(\frac{kb}{2}-\varphi_3)}  & 0 \\
	
	-\lambda e^{i(\frac{kb}{2}-\varphi_3)}  & 0 & 0  & \lambda 
e^{-i(\frac{kb}{2}-\varphi_3)} & 0  & 0 \\
	
	0 & 0 & 0  & 0 & 0  & 0 
\end{pmatrix}
\end{equation*}

\begin{equation*}
\bold{M}_4 =
	\begin{pmatrix}	
 \epsilon_1  & t e^{-i(kb_2+\varphi_2)} & 0  & t_c e^{ikb'} & t 
e^{-i(\frac{kb}{2}-\varphi_3)} & 0\\
	
 t  e^{i(kb_2 + \varphi_2)}  & \epsilon_2 & t_h  e^{-i(kb_1+\varphi_1)}  & 0 & 0 & 0\\
	
 0  & t_h  e^{i(kb_1+\varphi_1)} & \epsilon_2  & t e^{-i(kb_2+\varphi_2)} & 0 & 0\\
	
 t_c e^{-ikb'} & 0 & t  e^{i(kb_2 + \varphi_2)} & \epsilon_1 &  t 
e^{i(\frac{kb}{2}-\varphi_3)} & 0\\
	
 t e^{i(\frac{kb}{2}-\varphi_3)}  & 0 & 0  & t  e^{-i(\frac{kb}{2}-\varphi_3)} & \epsilon_3
& t_f \\
	
	 0  & 0 & 0  & 0 &  t_f  & \epsilon_4
	\end{pmatrix}
        \end{equation*}

Now the band structure can be deduced from the secular equation of 
the matrix $\bold{M}$, namely $\det(\bold{M}- \epsilon \bold{I})=0$, where 
$\epsilon$ represents the energy eigenvalues, while $I$ is the 
$12\times 12$ identity matrix. This leads to the following 
equation (see Appendix B):
\begin{eqnarray}
Q= \det( \bold{M}- \epsilon \bold{I})=C (A + i V)(A -i V) =0,
\label{equ11}
\end{eqnarray}
which represents in the present case Eq.(\ref{equ1}).
Here, $C= A_f^2  \bar{\epsilon}_2^2 \bar{\bar{\epsilon}}_2^2 \bar{\epsilon}_4^2 
\bar{\bar{\epsilon}}_3^2$,  $\bar{\bar{\epsilon}}_3 =\bar{\epsilon}_3 - 
\frac{|t_f|^2}{\bar{\epsilon}_4}, \bar{\bar{\epsilon}}_2 = \bar{\epsilon}_2 - 
\frac{|t_h|^2}{\bar{\epsilon}_2}$, $A_f = \bar{\epsilon}_1 - (t^2 + \lambda^2) 
\Big( \frac{\bar{\epsilon}_4}{\bar{\epsilon}_3 \bar{\epsilon}_4 - t_f^2} + 
\frac{\bar{\epsilon}_2}{\bar{\epsilon}_2^2 - t_h^2} \Big)$.
One has $\bar{\epsilon}_j=\epsilon_j - \epsilon$, ($j={1,2,3,4}$).
The expressions of $A$ and $V$ are detailed in Appendix B, and one has
\begin{eqnarray}
 &&(A + i V )= A_f -\frac{1}{A_f} \bigg(
(\bar{t}_c^* e^{i \varphi_k}
 - \frac{-\lambda^2 + t^2}{\bar{\bar{\epsilon}}_3} e^{i \varphi} 
 + \frac{-\lambda^2 + t^2}{\bar{\epsilon}_2 \bar{\bar{\epsilon}}_2} t_h) 
 + i (\bar{\lambda}_c e^{i\varphi_k} + \frac{2 \lambda t}{\bar{\bar{\epsilon}}_3} e^{i \varphi} - \frac{	2 \lambda t}{\bar{\epsilon}_2 \bar{\bar{\epsilon}}_2} t_h) 
	\bigg)
\nonumber\\
&&\hspace{1.4cm}
*\bigg( (\bar{t}_c e^{-i \varphi_k}
	-  \frac{-\lambda^2 + t^2}{\bar{\bar{\epsilon}}_3} e^{-i \varphi}
 + \frac{-\lambda^2 + t^2}{\bar{\epsilon}_2 \bar{\bar{\epsilon}}_2} t_h) - 
	i (\bar{\lambda}_c^* e^{-i \varphi_k} + \frac{2 \lambda t}{\bar{\bar{\epsilon}}_3} e^{-i \varphi} - \frac{	2 \lambda t}{\bar{\epsilon}_2 \bar{\bar{\epsilon}}_2} t_h) \bigg),
\nonumber\\
&&(A - i V)= A_f -\frac{1}{A_f} \bigg(
(\bar{t}_c^* e^{i \varphi_k}
- \frac{-\lambda^2 + t^2}{\bar{\bar{\epsilon}}_3} e^{i \varphi} 
+ \frac{-\lambda^2 + t^2}{\bar{\epsilon}_2 \bar{\bar{\epsilon}}_2} t_h) 
- i (\bar{\lambda}_c e^{i\varphi_k} + \frac{2 \lambda t}{\bar{\bar{\epsilon}}_3} e^{i \varphi} - \frac{	2 \lambda t}{\bar{\epsilon}_2 \bar{\bar{\epsilon}}_2} t_h) 
\bigg)
\nonumber\\
&&\hspace{1.4cm}
*\bigg( (\bar{t}_c e^{-i \varphi_k}
-  \frac{-\lambda^2 + t^2}{\bar{\bar{\epsilon}}_3} e^{-i \varphi}
+ \frac{-\lambda^2 + t^2}{\bar{\epsilon}_2 \bar{\bar{\epsilon}}_2} t_h) + 
i (\bar{\lambda}_c^* e^{-i \varphi_k} + \frac{2 \lambda t}{\bar{\bar{\epsilon}}_3}
e^{-i \varphi} - \frac{	2 \lambda t}{\bar{\epsilon}_2 \bar{\bar{\epsilon}}_2} 
t_h) \bigg),
\label{equ12}
\end{eqnarray}
where $ \bar{t}_c = t_c e^{2i \varphi_3}$,$\bar{\lambda}_c=\lambda_c e^{
-2i \varphi_3}$, $\varphi_k = ka +\varphi$, $\varphi$ =$\varphi_{1}$ + 
2 $\varphi_{2}$ + 2 $\varphi_3$ holds, and one has in 
Eq.(\ref{equ11}) the expression $Q=C I_{+}I_{-}=0$, $I_{\pm}= A \pm i V$, which 
cannot be satisfied by $C=0$.

In what will follow one analyzes the $I_{+}=0$ relation providing $Q=0$
(note that the same conclusions are provided by the $I_{-}=0$ relation, see
Appendix C). In the present situation, for $Q=0$ one has
\begin{eqnarray}
I_{+}= (A + i V)= T_0+ T_1  \cos(\varphi_k) + T_2 \sin(\varphi_k)=0,
\label{equ13}
\end{eqnarray}
where
\begin{eqnarray}
&&T_0=A_f -\frac{1}{A_f} \bigg( (\lambda_c^2 + t_c^2) + (\lambda^2 + t^2)^2 
\bigg( \frac{1}{\bar{\bar{\epsilon}}_3^2}+ \frac{t_h^2}{\bar{\epsilon}_2^2 
\bar{\bar{\epsilon}}_2^2}
- \frac{t_h}{\bar{\epsilon}_2 \bar{\bar{\epsilon}}_2 \bar{\bar{\epsilon}}_3} 2 
\cos( \varphi) \bigg) \bigg),
\nonumber\\
&&T_1= \frac{1}{A_f} \bigg(
-\cos(2 \varphi_3 + \varphi) \frac{2 (2 \lambda t \lambda_c  + t_c (\lambda^2 -
t^2))}{ \bar{\bar{\epsilon}}_3}
- \sin(2 \varphi_3 + \varphi) \frac{2 (-2 \lambda t t_c + \lambda_c (\lambda^2-
t^2))}{\bar{\bar{\epsilon}}_3} +
\nonumber\\
&&\hspace{2 cm} + \cos(2 \varphi_3) \frac{2 (2 \lambda  t \lambda_c + t_c (
\lambda^2 - t^2 )) t_h}{\bar{\epsilon}_2 \bar{\bar{\epsilon}}_2} 
+ \sin(2 \varphi_3) \frac{2 (-2 \lambda t t_c + \lambda_c (\lambda^2 - t^2)) 
t_h}{\bar{\epsilon}_2 \bar{\bar{\epsilon}}_2}\bigg),
\nonumber\\
&&T_2=\frac{1}{A_f}\bigg( 
\cos(2 \varphi_3 + \varphi) \frac{2 (-2 \lambda t t_c + \lambda_c (\lambda^2 - 
t^2))}{ \bar{\bar{\epsilon}}_3}
-\sin(2 \varphi_3 + \varphi) \frac{2 (2 \lambda  t \lambda_c + t_c (\lambda^2 -
t^2 ))}{ \bar{\bar{\epsilon}}_3}-
\nonumber\\
&&\hspace{2 cm} - \cos(2 \varphi_3)  \frac{2 (-2 \lambda t t_c + \lambda_c (
\lambda^2 - t^2)) t_h}{\bar{\epsilon}_2 \bar{\bar{\epsilon}}_2} 
- \sin(2 \varphi_3) \frac{2 (2 \lambda  t \lambda_c + t_c (\lambda^2 - t^2)) 
t_h}{\bar{\epsilon}_2 \bar{\bar{\epsilon}}_2} \bigg).
\label{equ14}
\end{eqnarray}
The here obtained $T_j=T_j({p_i})$,  $j=1,2$ are the terms present in 
Eq.(\ref{equ2}), and in the present case 
$trig_1({\bf k}{\bf x}_{\alpha})= \cos(\varphi_k)$, 
$trig_2({\bf k}{\bf x}_{\alpha})=\sin(\varphi_k)$ holds.
The flat band conditions become [see Eq.(\ref{equ2})]:
\begin{eqnarray}
T_1=0, \quad T_2=0.
\label{equ15}
\end{eqnarray}
From Eq.(\ref{equ13}) and the flat band conditions Eq.(\ref{equ15}) one also
has $T_0=0$. In general, this relations determines the position of the flat 
band.

\section{Relaxing the rigid flat band conditions while maintaining the 
position of the flat band}

\subsection{The rigidly fixed flat band conditions}

Let us start with the flat band conditions, Eq.(\ref{equ15}), in the absence of
SOI (i.e. $\lambda=\lambda_c=0$) and external magnetic field (i.e. $\varphi_i=0$
at $i=1,2,3$, see also Eq.(\ref{Aequ3}), i.e. $\phi=0$ as well). In doing this 
job we fix the origin of the energy axis to the position of the flat band
(i.e. $\epsilon=0$). From Eq.(\ref{equ15}) we find
\begin{eqnarray}
|t_{f}| = \frac{\sqrt{\epsilon_4[\epsilon_3 t_h
- (\epsilon_2^2 - t_h^2)]}}{\sqrt{t_h}},
\label{equ16}
\end{eqnarray}
while the $T_0=0$ condition, by fixing the flat band position to the origin,
provides
\begin{eqnarray}
|t_{c}| = \frac{(\epsilon_2 + t_h) (\epsilon_1 (\epsilon_2- t_h) - t^2)}{
\sqrt{(\epsilon_2^2 - t_h^2)^2}}.
\label{equ17}
\end{eqnarray}
These results are in agreement with the conditions deduced previously in
literature \cite{E20}. One notes that the sign of the ($t_f,t_c$) 
hopping amplitudes
influences the relative position of the flat band in the band structure of the
system. E.g. for ($t_f > 0, t_c > 0$) the flat band appears as the lowest band 
in the band structure [for $N_b$ number of atoms in the base (in our case 
$N_b=6$) one has $N_b$ 
bands in the band structure], for ($t_f < 0, t_c < 0$) the flat band appears in 
the upper position of the band structure, etc. 
From Eq.(\ref{equ16},\ref{equ17}) the meaning of rigid flat band conditions 
can be clearly
exemplified: All Hamiltonian parameters excepting $t_f,t_c$ can be arbitrarily
chosen (however the positivity conditions 
$\epsilon_4[\epsilon_3 -(\epsilon_2^2-t_h^2)/t_h] > 0, (\epsilon_2+t_h)[
\epsilon_1-(\epsilon_2-t_h)-t^2] > 0$ seen in Eqs.(\ref{equ16},\ref{equ17})
must be satisfied). But the $t_f$ value is rigidly fixed by Eq.(\ref{equ16}).
Furthermore, the Eq.(\ref{equ17}), by fixing the flat band position to 
$\epsilon=0$ fixes the $t_c$ value as well. In order to exemplify 
(see Set.1 of data in Appendix D), if one takes
e.g. $\epsilon_1 = 0.17, \epsilon_2 = 0.49, \epsilon_3 = 0.22, 
\epsilon_4 = 3.36, t_h = 1.5$ (as arbitrarily taken Hamiltonian parameters),
for the emergence of the flat band, we rigidly need $t_f=2.28941$, and in order 
to have $\epsilon=0$, we also need to have  ``rigidly'' $t_c=1.1601$. 
This rigidity is considered to be the main difficulty in obtaining the flat 
band in practice. One notes, that it often happens, that at $\epsilon=0$, the
rigid conditions relating the Hamiltonian parameters provided by $T_j =0, 
j > 1$ and $T_0=0$ become interdependent.
 
What we do now is as follows: maintaining the arbitrarily taken Hamiltonian 
parameters, we modify $t_f=t_f^{rfbc}$ value from the rigid condition 
Eq.(\ref{equ16}), and $t_c=t_c^{rfbc}$ value providing by Eq.(\ref{equ17}) 
leading to flat band at $\epsilon=0$, where $rfbc$ means ``rigid flat
band condition''. By this, the studied band becomes dispersive (details 
presented in Appendix D). But we show that now taking into account the 
SOI spin-orbit coupling, the relaxed $t_f = t_f^{rfbc}+ \Delta t_f$,
$t_c = t_c^{rfbc}+ \Delta t_c$ values are able
to provide a flat band again. Consequently, not only $t_f^{rfbc}$, $t_c^{rfbc}$
are able to provide the flat band, but also $t_f = t_f^{rfbc}+ \Delta t_f$,
$t_c = t_c^{rfbc}+ \Delta t_c$, do the same job,
hence the rigid flat band conditions can be relaxed by SOI. By this, taking
into account that $\lambda,\lambda_c$ can be tuned [even continuously e.g. by
an external electric field, see Eq.(\ref{equ6})], the set up of a flat band in 
practice becomes a more easier job. During this Section, in this process, by
keeping $\epsilon=0$, the flat band which emerges by re-flattening (i.e.
in the presence of  $\Delta t_f \ne 0$;  $\Delta t_c \ne 0$;
$\lambda, \lambda_c \ne 0$), will be placed at the origin of the energy
axis again. Hence here, we relax the rigid flat band conditions
but we maintain the position of the flat band at the same time. We do this job 
first in the absence of the $B$ external magnetic field. 

\subsection{Relaxing the rigid flat band conditions by SOI at $B=0$}

At $B=0$ and SOI present, based on Eq.(\ref{equ14}), the flat band condition
at $\epsilon=0$ presented in Eq.(\ref{equ15}) become
\begin{eqnarray}
&&T_1(B=0)= - \frac{2}{A_f}  \left( 2 
\lambda t \lambda_c  + t_c (\lambda^2 - t^2) \right) \left( \frac{\epsilon_4}{
\epsilon_{z}}  - \frac{t_h}{\epsilon_2^2 - t_h^2} \right) =0,
\nonumber\\
&&T_2(B=0) = - \frac{2}{A_f}  \left( 2 
\lambda t t_c  - \lambda_c (\lambda^2 - t^2) \right) \left( \frac{\epsilon_4}{
\epsilon_{z} }  - \frac{t_h}{\epsilon_2^2 - t_h^2} \right)=0,
\label{equ18}
\end{eqnarray}
while the $T_0=0$ relation maintaining the flat band in the origin provides
\begin{eqnarray}
T_0(B=0)=A_f -\frac{1}{A_f} \bigg( 
(\lambda_c^2 + t_c^2) + (\lambda^2 + t^2)^2 
\bigg( \frac{1}{\bar{\bar{\epsilon}}_3^2}+ \frac{t_h^2}{\bar{\epsilon}_2^2 
\bar{\bar{\epsilon}}_2^2}
-2  \frac{ t_h}{\bar{\epsilon}_2 \bar{\bar{\epsilon}}_2 \bar{\bar{\epsilon}}_3}
\bigg) \bigg) =0.
\label{equ19}
\end{eqnarray}
One notes, that the notations $A_f,\bar{\epsilon}_j, 
\bar{\bar{\epsilon}}_j,$ are given below in Eq.(\ref{equ11}). Furthermore,
because of $\epsilon=0$ one has $\bar{\epsilon}_j=\epsilon_j$,
and $\epsilon_{z}=\epsilon_3 \epsilon_4 - t_f^2$ holds.

In the two lines of Eq.(\ref{equ18}) the simultaneous zero value of the two
brackets containing $\lambda,\lambda_c$ 
requires not allowed complex SOI coupling values.
Hence Eq.(\ref{equ18}) is satisfied only by 
$\epsilon_4(\epsilon_2^2-t_h^2)=\epsilon_z t_h$, which leads to the 
$t_f=t_f^{rfbc}$ value presented in Eq.(\ref{equ16}). Consequently, if
we would like to maintain the position of the flat band (i.e. $\epsilon=0$
has been fixed), $t_f^{rfbc}$ cannot be relaxed by SOI couplings. But
$t_c^{rfbc}$ [presented in Eq.(\ref{equ17})] can be relaxed by SOI couplings.
In order to see this, first one
modifies $t_c$ to the value $t_c=t_c^{rfbc} + \Delta t_c$ and makes the flat band
dispersive. What is happening explicitly in this step is presented in details 
in Appendix E, where the dispersive band obtained from the flat band at 
$\Delta t_c \ne 0$ and missing SOI is characterized (e.g. see Fig.12).

In the second step we turn on the SOI, which according to Eq.(\ref{equ19}) is
able to turn the dispersive band - obtained in the first step - back to a flat 
band at the same position on the energy scale.

\begin{figure}[!htb]
	\includegraphics[width=\textwidth]{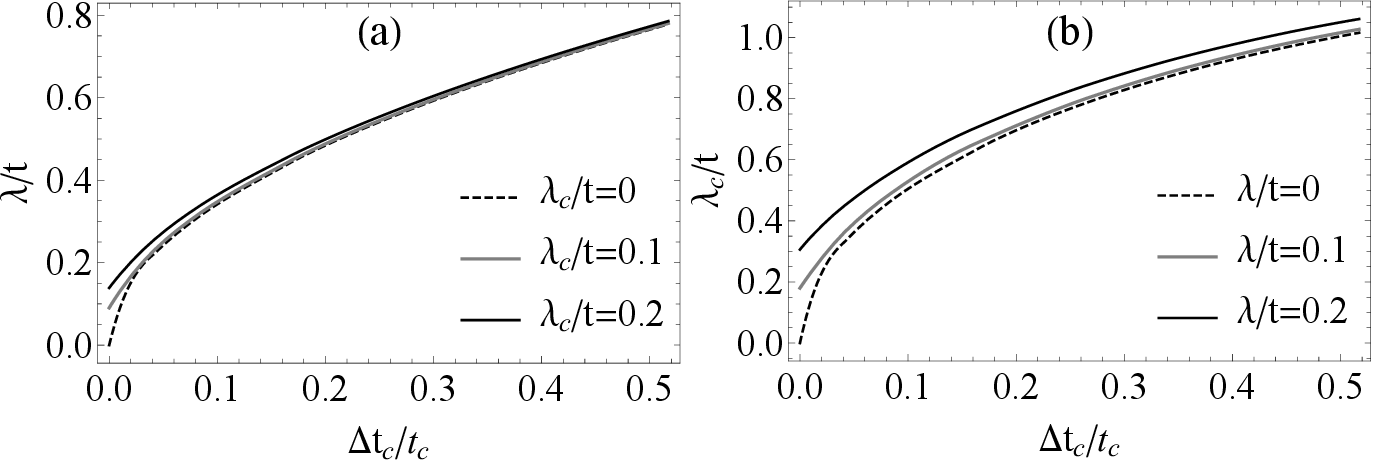}
	\caption{  (a) The $\lambda$ values necessary to achieve a flat band 
with different fixed $\lambda_c$ values, after changing the $t_c=t_c^{rfbc}$ 
value defined by the rigid flat band conditions by $\Delta t_c$ (b) The 
$\lambda_c$ values necessary to achieve a flat band with different fixed 
$\lambda$ values, after changing the $t_c=t_c^{rfbc}$ value defined by the rigid
flat band conditions by $\Delta t_c$. For exemplification we have used the
Set.1 of Hamiltonian parameter data from Appendix D.}
	\label{fig8}
\end{figure}

\begin{figure}[!htb]
	\includegraphics[width=\textwidth]{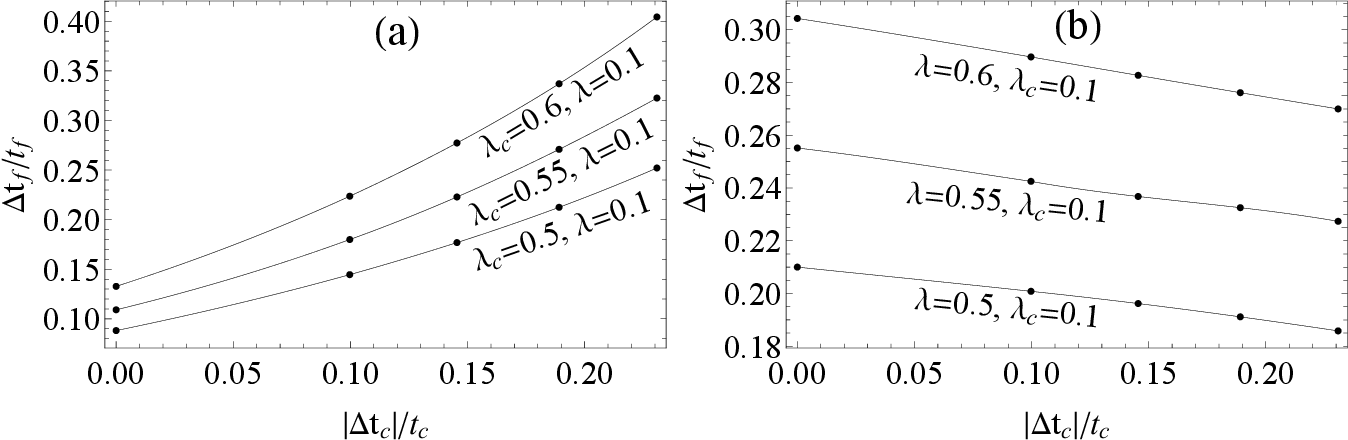}
	\caption{  The $\lambda$,  $\lambda_c$ values necessary to achieve a 
flat band after changing the $t_c=t_c^{rfbc}$ value defined by the rigid
flat band conditions by $\Delta t_c$. The changed $\Delta t_f$ values were 
obtained by changing the $t_h$ hopping magnitude, therefore the value of 
$t_h$ is continuously changing along the solid lines, form the original $t_h$ 
to $1.5 t_h$. In case a) one has $\lambda=constant$, while in the case b)
$\lambda_c=constant$ holds. For exemplification we have used the
Set.1 of Hamiltonian parameter data from Appendix D.}
	\label{fig9}
\end{figure}
What one obtains is exemplified in Fig.4. The arbitrarily chosen Hamiltonian 
parameters, together with $t_f^{rfbc}$ and $t_c^{rfbc}$ are those used in 
Appendix D. One observes that even $50 \%$ change in $t_c^{rfbc}$ can be easily
compensated by $\lambda$ or $\lambda_c$ in reproducing the flat band in its
initial position. One further observes that the in-base SOI ($\lambda$) is
more efficient than its inter-base ($\lambda_c$) counterpart, since smaller
$\lambda$ values are able to compensate the same $\Delta t_c$ values in 
reproducing the flat band. As seen, indeed the rigid flat band condition is
substantially relaxed by SOI, at least at the level of $t_c$. 
The price of the re-flattened band to remain in the same position is that 
not all rigidly fixed Hamiltonian parameters can be relaxed (such as 
$t_f^{rfbc}$ in the present case). In such situation traditional procedures
can be combined with SOI in order to achieve the re-flattening after the
application of the $\Delta t_c \ne 0$ rigid flat band condition relaxation.
In this case $\Delta t_f$ can be obtained by changing the side group connected
to the pentagon [see Fig.2, where the side group $N H_2$ appears in the top
(apical) part of the figure]. In the same time with this step, the counter
apical (i.e. the N-N bond in Fig.2) hopping matrix element $t_h$ must be 
modified, which can be achieved e.g. by doping polyaminotriazole with $ClO_4$ 
($PAT ClO_4$), fluorine ($PAT F$), $HF_2$ ($PAT HF_2$), etc. What is obtained 
is exemplified in Fig.5. We must here underline that higher $\lambda_c$
values allow higher $\Delta t_f$ values to be achieved in the  attempt to 
transform the band back to a flat band. On this line we mention that by
introducing heavy ions on intercell bonds we are able to increase the SOI 
coupling along the inter-base bonds \cite{E57}, and as seen here, this step
would allow to increase the deviation from ``rfbc'' values in the process of
band flattening.

\begin{figure}
        \includegraphics[width=7.5cm]{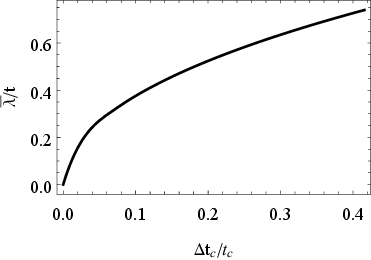}
	\caption{The $\bar \lambda = \lambda = \lambda_c$ value necessary to
compensate the deviation $ \Delta t_c \ne 0$ from 
$t_c^{rfbc}$ in order to re-create the flat band at the origin of the energy 
axis. For exemplification we 
have used the Set.1 of Hamiltonian parameter data from Appendix D.}
	\label{fig6}
\end{figure}

Often it happens that one has a pentagon polymer chain in which external side 
groups (apical atoms) are not present, doping is not used, and also heavy ion 
introductions on inter-base bonds is missing. 
In this case $\lambda=\lambda_c=\bar \lambda$ values can
be tuned by external electric field as specified in Eq.(\ref{equ6}). In such
conditions, deviations $\Delta t_c \ne 0$ from $t_c^{rfbc}$ can be compensated by
$\bar \lambda$ as shown in Fig.6 in order to create back the flat band at the 
origin of the energy axis. 

\subsection{Relaxing the rigid flat band conditions by SOI at $B \ne 0$}

When $B \ne 0$ holds, the flat band conditions Eq.(\ref{equ15}) 
can be written as
\begin{eqnarray}
\frac{1}{A_f} (-K_g v - S_g u)=0, \quad \frac{1}{A_f} (K_g u - S_g v )=0,
\label{equ20}
\end{eqnarray}
where the following notations have been introduced
\begin{eqnarray}
&&K_g=\frac{\cos(4 \varphi_3 + \varphi_b)}{ \bar{\bar{\epsilon}}_3} - 
t_h \frac{\cos(2 \varphi_3)}{\bar{\epsilon}_2 \bar{\bar{\epsilon}}_2}, \quad
S_g=\frac{\sin(4 \varphi_3 + \varphi_b)}{ \bar{\bar{\epsilon}}_3} - 
t_h \frac{\sin(2 \varphi_3)}{\bar{\epsilon}_2 \bar{\bar{\epsilon}}_2},
\nonumber\\
&&v= 2 (2 \lambda \lambda_c t + t_c (\lambda^2 - t^2 )), \quad
u=2 (-2 \lambda t t_c + \lambda_c (\lambda^2 - t^2)),
\nonumber\\
&&\varphi= \varphi_b + 2 \varphi_{3}, \quad
\varphi_b =\varphi_1 + 2 \varphi_{2}.
\label{equ21}
\end{eqnarray}
Since for the Rashba interaction considered here $\lambda, \lambda_c$ must 
be real,
Eq.(\ref{equ20}) allows solutions only for $K_g=S_g=0$, which provide 
\begin{eqnarray}
&&I_c=I_s=I_{\varphi}, \quad I_{\varphi}=X_{\varphi}, 
\nonumber\\
&&I_c=\frac{\cos(2\varphi_3)}{\cos(4 \varphi_3+ \varphi_b)}, \quad
I_s=\frac{\sin(2\varphi_3)}{\sin(4\varphi_3+\varphi_b)}, \quad
X_{\varphi}=\frac{\epsilon_4}{t_h} \frac{(\epsilon_2^2 - t_h^2)}{
(\epsilon_3 \epsilon_4 - t_f^2)}. 
\label{equ22}
\end{eqnarray}
For solving Eq.(\ref{equ22}) one studies the equality $I_s=I_c$. 
Before starting this job, let us underline that
in the limit of zero external magnetic field, this equality gives
$I_{\varphi}=X_{\varphi}=1$ and we reobtain the $B=0$ flat band condition 
deduced previously
in Eq.(\ref{equ16}). For $B \ne 0$, using $\sin(\alpha-\beta)=\sin(\alpha)
\cos(\beta)-\cos(\alpha)\sin(\beta)$, the $I_s=I_c$ relation gives
\begin{eqnarray}
\sin[(4\varphi_3+\varphi_b)-2\varphi_3]=
\sin (\varphi) = 0, \quad i.e. \quad \varphi= \pm n \pi,
\label{equ23}
\end{eqnarray}

\begin{figure}
	\includegraphics[width=15cm]{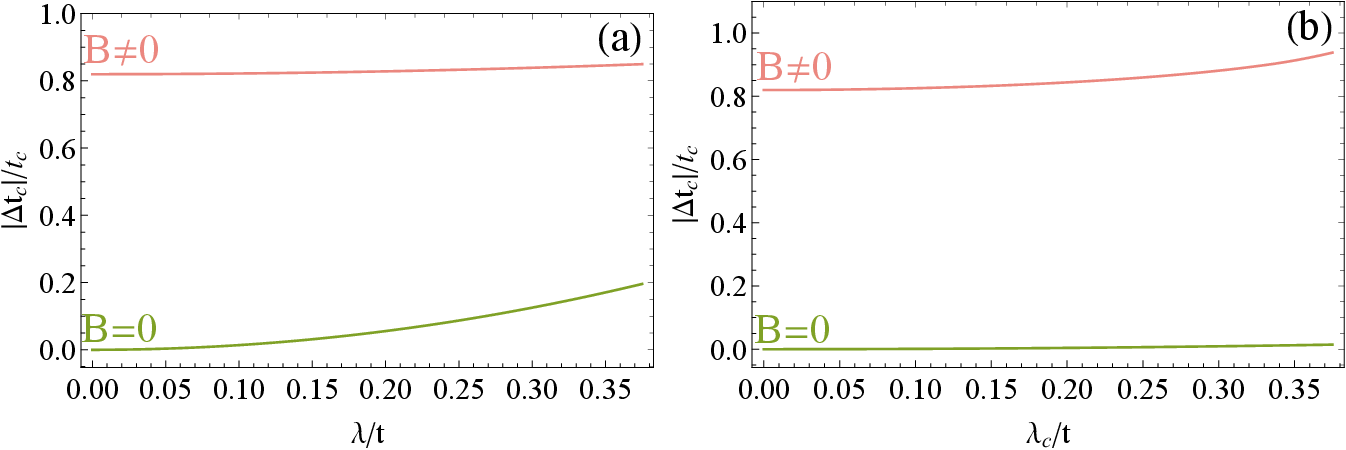}
	\caption{a) $\lambda$ and b) $\lambda_c$ spin orbit coupling 
values deduced at $B \ne 0$ (upper) and $B = 0$ (lower) necessary to
compensate the deviation $\Delta t_c \ne 0$ from $t_c^{rfbc}$ in order to 
re-create the flat band at the origin of the energy axis. 
For exemplification we 
have used the Set.2 of Hamiltonian parameter data from Appendix D.}
	\label{fig7}
\end{figure}
%
%
%
\begin{figure}
\includegraphics[width=7.3cm]{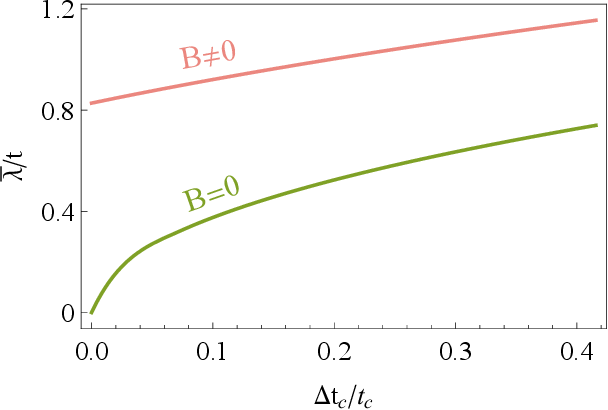}
	\label{fig8}
	\caption{The $\bar \lambda=\lambda=\lambda_c$ spin orbit coupling 
		values deduced at $B \ne 0$ (upper) and $B = 0$ (lower)
necessary to compensate the deviation $\Delta t_c \ne 0$ from $t_c^{rfbc}$ 
in order to re-create the flat band at the origin of the energy axis.
For exemplification we have used the Set.3 of Hamiltonian parameter data 
from Appendix D.}
\end{figure}

where $n$ is an integer number or zero. Hence at $I_s=I_c$, one obtains
$I_s=\sin(2\varphi_3)/\sin(\pm n \pi + 2\varphi_3)$, consequently
$I_{\varphi}=X_{\varphi}=\pm 1$, where the upper sign is obtained 
at $n=0$, while the
lower sign at other $n$ values. One further observes that when Eq.(\ref{equ23})
holds, Eq.(\ref{equ19}) remains true, so the first line of Eq.(\ref{equ14})
reduces to  Eq.(\ref{equ19}) when the flat band appears in the presence of the
external magnetic field in the same position of the energy axis. Since 
$X_{\varphi}=1$, as mentioned above, reproduces the $B=0$ results (i.e. 
$\Delta t_c=\Delta t_f=0$ holds in this case), the $B \ne 0$ characteristics can
be derived from the $X_{\varphi}=-1$ relation. Based on the last equality of the
second line of Eq.(\ref{equ22}), we obtain four different possible deviations 
$\Delta t_f$ from the $t_f^{rfbc}$ value, which are able to re-flatten the band
at $B \ne 0$ in the same position of the energy axis in which it was placed
the flat band at $B=0$:
\begin{eqnarray}
\Delta t_f = \pm \frac{\sqrt{\epsilon_4[\epsilon_3 t_h
- (\epsilon_2^2 - t_h^2)]}}{\sqrt{t_h}} \pm \frac{\sqrt{\epsilon_4[\epsilon_3 t_h
+ (\epsilon_2^2 - t_h^2)]}}{\sqrt{t_h}},
\label{equ24}
\end{eqnarray}
How $\Delta t_c$  modifies as function of the spin orbit coupling 
in flattening the
band at $B \ne 0$ (placing the flat band in the same position of the energy 
axis in which the flat band for $t_f=t_f^{rfbc}, t_c=t_c^{rfbc}$ was placed
at $B=0$) is exemplified in Fig.7.  The presented $|\Delta t_c|/t_c$ results 
were deduced from the $T_0$ expression of Eq.(\ref{equ14}) in condition of
Eq.(\ref{equ23}). The $\bar \lambda=\lambda=\lambda_c$ results are similar, and
are presented in Fig.8.

Based on the results presented in this subsection relating the $B \ne 0$ case,
the following observations can be made:
1) It can be observed that only discrete nonzero external magnetic field values
provide re-flattening effects [see Eq.(\ref{equ23})].
2) As shown by Fig.7 and Eq.(\ref{equ24}), huge $\Delta t_c/t_c$ and
$\Delta t_f/t_f$ values can be achieved at $B \ne 0$ [allowed by the point 1)]
in relaxing the rigid flat band conditions necessary for obtaining a flat band
in the same position of the energy axis. Fig.7 shows that 80\% deviations from
$t_c^{rfbc}$ can be easily compensated by relatively small spin-orbit 
interaction 
values, and based on Eq.(\ref{equ24}) it can be checked that 40-50\% deviations
from $t_f^{rfbc}$ can be achieved in producing a flat band at nonzero $B$. 
3) The requirement to maintain a fixed flat band position on the energy axis
is relatively restrictive since it does not allow all rigidly fixed flat band 
conditions to be continuously relaxed. In the present case, at $B \ne 0$, the
$t_f^{rfbc}$ can be only discretely modified when flat bands are intended to be
manufactured. 4) Since the condition in Eq.(\ref{equ23}) is connected only to
the total flux threading the unit cell, it results that
in distorting the unit 
cell, new aspects in the band flattening via spin orbit interaction are not 
encountered. 

\section{Relaxing the rigid flat band conditions without maintaining the 
position of the flat band}

Let us consider that one has a flat band at $B=\lambda=\lambda_c=0$ which is
placed in the origin of the energy axis, i.e. at $\epsilon=\epsilon_1=0$. As 
described previously, for the Hamiltonian parameters, this flat band emergence 
requires rigidly fixed flat band conditions,
e.g. in the present case $t_f=t^{rfbc}_f, t_c=t^{rfbc}_c$. Now we modify $t_f$ and
$t_c$ with $\Delta t_f$ and $\Delta t_c$ relative to $t_f=t^{rfbc}_f$, and
$t_c=t^{rfbc}_c$, the studied flat band becoming dispersive as exemplified in
Appendix E. After this step we turn on the SOI such to transform back the
dispersive band obtained in the previous step into a flat band placed in the
position $\epsilon=\epsilon_2$. In the previous Section IV., we have analyzed
the characteristics of this re-flattening process for $\epsilon_1=\epsilon_2=0$,
i.e. for the case in which the re-flattened band emerges in the same position of
the energy axis. Contrary to this, in the present Section V. we will analyze
the case  $\epsilon_1 \ne \epsilon_2$, i.e. the situation in which the starting
flat band position $\epsilon_1$ obtained at  $t_f=t^{rfbc}_f, t_c=t^{rfbc}_c$, and 
$B=\lambda=\lambda_c=0$, will be different from the position $\epsilon_2$ 
of the flat band obtained via SOI at the end of the process. As it will be seen
from the results, this situation allows to considerably relax all rigidly 
fixed flat band conditions, hence allows to manufacture flat bands in real 
systems under easier conditions.   
\begin{figure}
	\includegraphics[width=17cm]{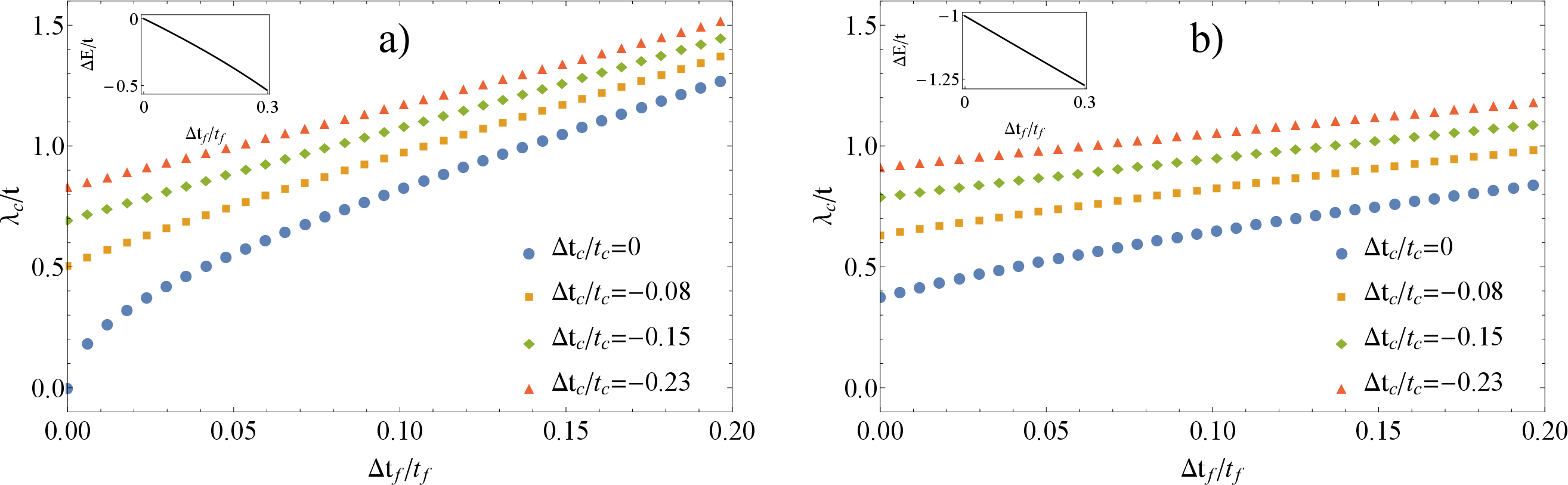}
	\caption{The $\lambda_c$ spin orbit coupling 
values necessary to compensate at $\lambda=0$ the common deviations
$\Delta t_c \ne 0$ 
and $\Delta t_f \ne 0$ at  a) $B=0$, and b) $B \ne 0$, in order to re-create a
flat band placed originally (at $\lambda=\lambda_c=B=0$) in the origin of the
energy axis $\epsilon=0$. The new position of the flat band is at
$\epsilon=\Delta E$. In this figure $\Delta t_c/t_c < 0$ holds.
For exemplification we have used the Set.4 of Hamiltonian 
parameter data from Appendix D.}
	\label{fig9}
\end{figure}

\begin{figure}
	\includegraphics[width=17cm]{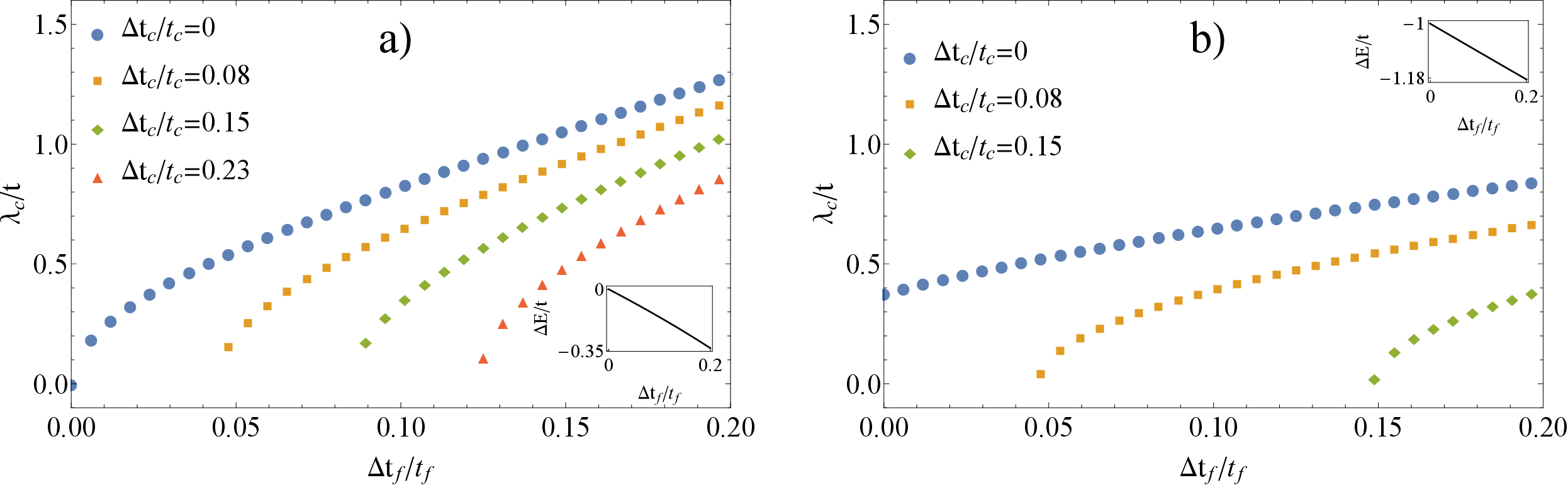}
	\caption{The $\lambda_c$ spin orbit coupling 
values necessary to compensate at $\lambda=0$
the common deviations $\Delta t_c \ne 0$ 
and $\Delta t_f \ne 0$ at a) $B=0$, and b) $B \ne 0$ in order to re-create a
flat band placed originally (at $\lambda=\lambda_c=B=0$) in the origin of the
energy axis $\epsilon=0$. The new position of the flat band is at
$\epsilon=\Delta E$. In this figure $\Delta t_c/t_c > 0$ holds.
For exemplification we have used the Set.4 of Hamiltonian 
parameter data from Appendix D.}
	\label{fig10}
\end{figure}

\begin{figure}
	\includegraphics[width=17cm]{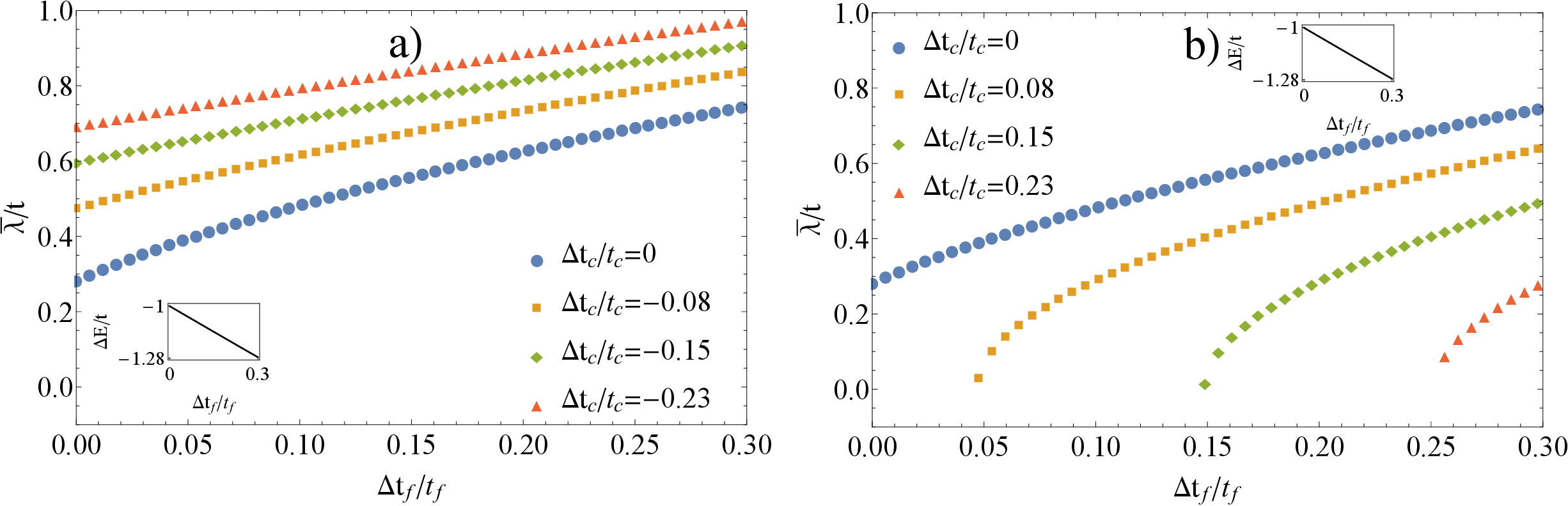}
	\caption{The $\lambda_c=\lambda=\bar \lambda$ spin orbit coupling 
values necessary to compensate the common deviations 
$\Delta t_c \ne 0$ and $\Delta t_f \ne 0$ at $B \ne 0$ and 
a) $\Delta t_c/t_c < 0$, and b) $\Delta t_c/t_c > 0$, in order to re-create a
flat band placed originally (at $\bar \lambda=B=0$) at the origin of the
energy axis $\epsilon=0$. The new position of the flat band is at
$\epsilon=\Delta E$. For exemplification we have used the Set.4 of Hamiltonian 
parameter data from Appendix D.}
	\label{fig11}
\end{figure}

Fig.9 (at $\Delta t_c/t_c < 0$) and Fig.10 (at $\Delta t_c/t_c > 0$) 
exemplifies the obtained results at $\lambda=0$. In these figures, the a) 
plots show the $B=0$ case, while the b) plots the $B \ne 0$ situation. It can 
be seen that even 20 $\%$ modification of $t_c^{rfbc}$ or $t_f^{rfbc}$ can be 
compensated by the presence of $\lambda_c$ in reproducing the flat band in a
shifted position $\Delta E$ presented in the inset. It can be observed that all
rigidly fixed Hamiltonian parameters can be relaxed in this case. In the 
presence of the external magnetic field $B$ larger $\Delta t_f/t_f$ deviations
can be compensated by smaller $\lambda_c$ values, which underlines the 
importance of the consideration of $B$ in this process. In this case $\lambda_c$
can be modified by atom intercalation in the intercell bonds, structural 
conformation or twist application \cite{E51,E53,E54a,E54b}.

When one considers the intrinsic $\lambda$ and $\lambda_c$ small, and we tune
both of them by external electric field, the $\lambda=\lambda_c=\bar \lambda$
case must be considered presented in Fig.11 which is plotted at nonzero and
constant $B$. It can be observed that e.g. almost 30 $\%$ positive displacements
in $\Delta t_c/t_c$ and $\Delta t_f/t_f$ can be compensated by relatively small
$\bar \lambda/t$ values of order $10^{-2}$.

\section{Further remarks}

Several observations and remarks we would like to add in what will follows.

a) We approach the presented subject in fact in a mathematical language
since this allows to show how the flat bands
can be detected in general terms in an arbitrary case, how the flat band 
conditions can be deduced in a general case, shows that a part of the flat 
band conditions enumerated in the literature in fact provide the position of 
the flat band on the energy scale.

b) We have used different input parameters in Table I-II presented in
Appendix D in order to underline that our findings, observations and technical
approach to the problem are not related only to one given material, but have 
an extremely broad application spectrum. The used Hamiltonian parameters are 
not new, have not been introduced in this paper, all of them have a broad 
literature. Consequently, how these parameters affect the band structure 
is known. For example, in the case of conducting polymers, how the hopping 
parameters ($t_i$) and on-site one particle potentials ($\epsilon_i$) influence
the band structure is seen e.g. in \cite{E18,E20,EX1,EX2} etc. How the 
Peierls phase factors -- describing the action of the external magnetic field 
on the orbital motion of itinerant electrons -- acts on the band structure is 
seen e.g. in \cite{E21,EX3,EX4} etc.  How the many-body spin-orbit
interaction acts on the band structure (in most cases the main effect is that 
breaks the spin projection double degeneracy of each band), is seen e.g. in
\cite{E38,E39,EX5,EX6}, etc.
This is why in this paper we concentrate on a single band not satisfying, 
but being in the absence of spin-orbit interaction closely placed to the rigid 
flat band conditions.

c) In order to exemplify inside the whole band structure the many-body SOI 
flattening effect we present Fig.12, where the polyaminotriazole case is 
exemplified in zero external magnetic field. The used Hamiltonian parameters
are $t_f/t =1$ (see Ref.\cite{E18,EX1}), and $t_h/t=0.93, t_c/t=1.06,
\epsilon_1/t=\epsilon_2/t=0.33, \epsilon_3/t=1.66, \epsilon_4/t=0.8$, see
Ref.\cite{EX7}. The rigid flat band conditions require in the absence of
external fields and $\lambda=\lambda_c=0$ the values $t_f^{rfbc}=1.406$, 
$t_c^{rfbc}=1.996$ (in $t$ units). As seen, $\lambda/t=0.17$ provides
$\Delta t_f/t_f=0.40$, $\Delta t_c/t_c=0.88$ relaxation of rigid flat band 
conditions.
\begin{figure}
	\includegraphics[width=15cm]{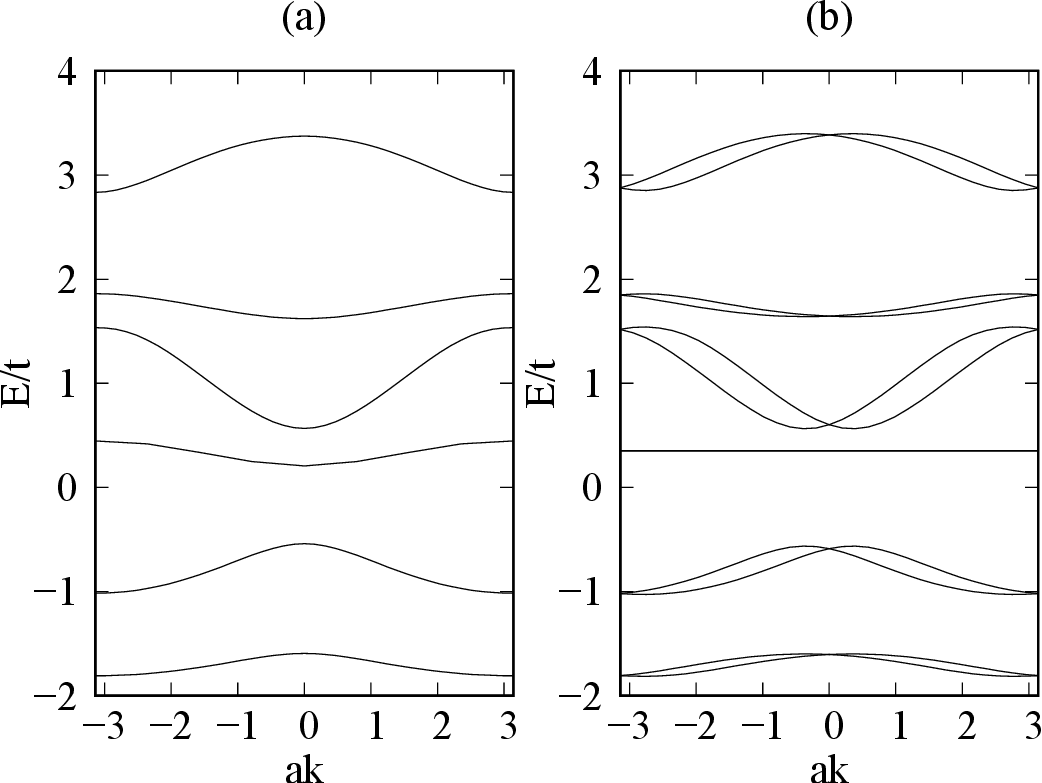}
	\caption{The many-body spin-orbit interaction effect on the whole
band structure: a) the $\lambda=\lambda_c=B=0$ case providing dispersive bands,
and b) at $B=0$ the $\lambda/t=0.178$ providing a double degenerate flat band
(see Appendix C). For the Hamiltonian parameters see text.}
	\label{fig12}
\end{figure}
In the presented case, the flat band emerges at the position of the third band
(see Fig.12.a). One notes, that in general, sign changes in the Hamiltonian 
parameters change the position of the resulting flat band.

d) Concerning the contribution of terms not taken into account in the 
Hamiltonian presented in Eq.(3) the following aspects must be underlined:
First, as mentioned previously in the text following Eq.(5), in the used
configuration the Zeeman term provides zero contribution because when the
external magnetic field ${\bf B}$ is applied, the carrier spin is perpendicular
to the magnetic field, hence the scalar product ${\bf B} \cdot {\bf \sigma}=0$
holds.  Second, if electric dipole moment ${\bf p}$ is present, it provides a 
supplementary contribution to the applied external electric field. In the case 
of the described conducting polymers, the dipole moment vector (if exists) is 
placed inside the plane x0y of the polymer. The dipole moment itself 
originates usually from the inside of the unit cell, being relatively small,
as order of magnitude around or below 1 debye \cite{EX8}. Our analyzed external 
electric field ${\bf E}$ is perpendicular to the plane of the polymer, so 
${\bf E}\cdot {\bf p}=0$, consequently the Stark contribution is also missing
from the Hamiltonian.

But it must be mentioned, that in special cases, in order to enhance special
applications (e.g. in energy storage, or solar cell manufacturing), it is 
possible to attach to the polymer \cite{EX9} group of atoms with high dipole 
moment, even oriented outside of the polymer plane. In such cases the presented
results and findings remain true, but the external electric field is additively 
renormalized by the electric field created by the dipole moments.  

e) If we concentrate on the question why the flat bands emerge, the answer to
this question underlined in this paper is as follows:
The band structure is given by the secular equation [Q=0, see Eq.(1)]
of the one particle part of the Hamiltonian transformed in the k-space. 
As explained, always  
\begin{eqnarray}
Q= T_0 + \sum_{j=1}^m T_j \: trig_j ({\bf k}{\bf x}_{\alpha}),
\label{equ25} 
\end{eqnarray}
where ${\bf x}_{\alpha}$ are the Bravais vectors, ${\bf k}$ the momentum, and 
the prefactors $T_0,T_1,...T_m$ depend only on the Hamiltonian 
parameters $\{p_i\}$ and the energy $\epsilon$. The bands are given by 
$\epsilon=E_n({\bf k})$ solutions of the $Q=0$ secular equations. Flat bands 
are obtained always when all $T_j=0$, see Eq.(2), hence $\epsilon$ becomes 
${\bf k}$ independent (and the flat band position will be given by the $T_0=0$ 
relation). This is the mathematical origin of the flat bands. How we 
achieve the all $T_j=0$ requirements ? We simply tune the Hamiltonian 
parameters $\{p_i\}$. We underline that the here described procedure can be 
applied always. It also means that flat bands are not the privilege of some 
special systems, since flat bands can be obtained from each system by a proper 
tuning of the Hamiltonian parameters, procedure which is effectively often used 
(see e.g. Refs. \cite{EX10,EX11}). E.g. for a simple cubic lattice in a simple 
tight binding approximation, from Q=0 one obtains for the lowest band the 
relation $\epsilon-A_0 - A_1 [cos(x k_x) + cos(y k_y) + cos(z k_z)]=0$,
where $T_0=\epsilon-A_0, T_1=A_1 \sim t$ holds, $t$ being the nearest 
neighbor hopping integral. The $T_1=A_1=0$ condition provides a flat 
band at the position $\epsilon=A_0$.  

Now if we ask: what it happens physically when all $T_j=0$ relations hold,
the answer to this question varies from case to case. For example, in the 
simple cubic lattice case exemplified above, $A_1=0$ occurs when 
the nearest-neighbor overlap becomes zero (e.g. when we increase the lattice 
constant at fixed itinerant carrier number, i.e. decrease the carrier 
concentration), and we reach a low concentration insulating localized state,
(e.g. Wigner lattice, since the Coulomb repulsion is always present). 
The problem becomes complicated also because there are flat bands with 
itinerant (i.e. non-localized) carriers (see e.g. Ref. \cite{E13}). In the 
situation when carriers are localized in the flat band, often the all $T_j=0$
relations are considered related to destructive interference caused by 
frustration, or lattice geometry (see e.g. Refs. \cite{EX10,EX11}).
In the conducting polymer case exemplified in this paper, one knows that in the
flat band, the one-particle Wannier states are extended over two cells but are 
localized (see e.g. Fig.1 of Ref. \cite{EX1}). These Wannier states
can be expressed as a linear combination of extended Bloch states. Hence, in 
explaining the two cell extension of the Wannier states, the 
destructive interference argument can be invoked also here. 

It is important to stress, that independent on how we interpret physically
Eq.(2), i.e. the all $T_j=0$ relations, the here described procedure in 
deducing the flat band emergence always works.

\section{Summary and conclusions}

One knows that in a real system, the emergence of a flat band is connected to
rigid mathematical conditions (i.e. flat band conditions) relating a part of 
the Hamiltonian parameters (which we denote here by $\xi_i$, e.g. in the 
presented paper, $\xi_1=t_c, \xi_2=t_f$).
Because of these rigid and restrictive conditions, the engineering
of a flat band in a real system is a quite difficult task. Indeed, for this to 
be possible, the rigidly fixed Hamiltonian parameters must be tuned exactly to 
the values fixed by the flat band conditions in order to obtain a flat band in
the system. From the other side, given by their high (practically infinitely
large) degeneracy, there is a huge need for flat bands in different systems,
because introducing a small perturbation in such case, the ground state of
such materials can be easily pushed in the direction of several ordered phases
of interest in different applications. Because of these reasons, the study
of procedures that are able to relax the rigid flat band conditions is an
important task. 

On this line, in this paper we demonstrate, that the many-body spin-orbit 
interaction (SOI) is able to substantially relax the rigid flat band conditions,
and at the same time can be continuously tuned by external fields. Consequently
taking SOI into account, the flat band manufacturing in real systems becomes an
easier task.     

The problem detailed above is analyzed in the case of conducting polymers.
Besides the broad application possibilities of these materials, the motivation 
of this choice is the fact that the mathematical background of the flat band 
conditions can be presented in this case in full generality but in a 
clear, visible and understandable manner. One even has the possibility to 
analyze the action of in-cell ($\lambda$), and inter-cell ($\lambda_c$) SOI 
contributions separately. The procedure
we use is simple: first, fixing the position  of the flat band at 
the origin of the energy axis $\epsilon=0$, we deduce the flat band conditions 
at zero external fields and zero SOI. Then, for a 
fixed set of Hamiltonian parameters that can be arbitrarily chosen, we deduce 
the rigidly fixed values of Hamiltonian parameters $\xi_i=\xi_i^{rfbc}$.
After this step we destroy the flat band (transforming it into a dispersive 
band)
by modifying $\xi_i$ from $\xi_i^{rfbc}$ to $\xi_i'=\xi_i^{rfbc}+\Delta \xi_i$, 
and analyze what SOI values transform the dispersive band back into a  
flat band 
placed in the position $\epsilon'$. In this manner, at the appearance of the 
flat band, the parameters $\xi_i$ are no more rigidly fixed to $\xi_i^{rfbc}$,
but take the values $\xi_i'$, hence are relaxed by $\Delta \xi_i=\xi_i'-
\xi_i^{rfbc}$.

In the first step we analyze the case $\epsilon=\epsilon'$, so the destroyed
flat band, after the application of SOI arrives back in its original position.
This situation is usually considered in the literature, and is in fact
restrictive since it does not allow to relax all rigidly fixed flat band 
conditions.
The relaxed parameters however, calculated as $\Delta \xi_i/\xi_i$, 
can be easily changed by $20-30 \%$ relative to their initial value.
The application of an external $B$ magnetic field increases (at
fixed SOI) the possible  $\Delta \xi_i/\xi_i$ values even
to 80 $\%$ (see e.g. Fig.7).  Comparing to the case mentioned above, as a 
novelty, we also analyze the case $\epsilon' \ne \epsilon$ in the second step.
In this situation, in fact, mathematically, one of the flat band conditions is 
missing, so the rigid flat band conditions are not so restrictive. In this 
case, the relative flat band position displacement on an arbitrary scale 
$|\Delta \epsilon/\epsilon|$ is relatively small (i.e. 10-20 $\%$), and 
contrary to the first case, all rigidly fixed Hamilton operator parameters
$\xi_i^{rfbc}$ can be relaxed by 20-30 $\%$ with relatively small SOI coupling
values (see e.g. Fig.11.b, where even $\lambda/t < 0.1$). 
Also in this case, the presence of the external magnetic field enhances the 
relaxation process of the rigidly fixed Hamiltonian parameter values.

Concerning the question: how can the SOI couplings be modified and tuned, 
several
possibilities exist. One has discrete tuning possibilities, as for example
intercalation of elements with high spin-orbit coupling on bonds connecting
cells (e.g. intrachain heavy atoms), hence modifying $\lambda_c$. But more 
promising possibilities are 
provided by continuous modification possibilities as for example via 
torsioning, twisting, or application of external electric field. From these, 
the last possibility seems to be the most attractive (from the data published 
in the literature, see e.g. [\cite{E54}], $\lambda \sim 0.02$ eV is
attained usually by electric fields of order $E \sim $ kV/cm).

We strongly hope that the presented results will considerably enhance the
flat band engineering of real materials.


\appendix

\section{The Peierls phase factors}

In calculating the  $\varphi_{ji}$  Peierls phase factors, one follows Fig.1. 
In the
presence of the external magnetic field $B \ne 0$ they modify the hopping
matrix elements according to the relation 
\begin{eqnarray}
&& t_{j \leftarrow i}(B)= t_{j \leftarrow i}(0) 
e^{i \frac{2 \pi }{\phi_0} \int_i^j \vec{A} \vec{dl}} = t_{j \leftarrow i}(0) e^{i \varphi_{ji}} ,
\label{Aequ1}
\end{eqnarray}
where $\phi_0=\frac{hc}{e}$ is the flux quantum. One has 
$\varphi_{ji} = \frac{2 \pi }{\phi_0} 
\int_i^j \vec{A} \vec{dl}$, and  $t_{j \leftarrow i}(0)$ are the $B=0$ hopping 
matrix elements. Since $B=rot(\vec{A})$ holds, and $B$ points to the $z$ 
direction, we use the gauge $A_x = -By, A_y=A_z=0$. After this step all
exact $\varphi_{ji}$ Peierls phase factors can be calculated for each bond.
One observes that $\varphi_{56}=0$, since the scalar product is zero 
($\vec{A} \perp \vec{dl} $), and $\varphi_{47}$ is also $0$, because $y=0$ 
holds (see Fig.\ref{fig1}). One obtains
\begin{eqnarray}
&& \varphi_{3 \leftarrow 2}= \frac{2 \pi }{\phi_0} (-By_2), \varphi_{3 \leftarrow 2} 
= \varphi_{1},
\nonumber\\
&& \varphi_{4 \leftarrow 3} = \frac{2 \pi B }{\phi_0} \frac{|y_2| b_2}{2}, 
\varphi_{2 \leftarrow 1} = \varphi_{4 \leftarrow 3} = \varphi_{2},
\nonumber\\
&&  \varphi_{5 \leftarrow 4} = \frac{2 \pi }{\phi_0}  B \frac{y_1 b}{4}, \quad  
\varphi_{5 \leftarrow 4} = \varphi_{1 \leftarrow 5} = \varphi_{3}.
\label{Aequ2}
\end{eqnarray}
One further has
\begin{eqnarray}
\varphi = \varphi_1+2\varphi_2+2\varphi_3= \frac{2\pi}{\phi_0} \phi,
\label{Aequ3}
\end{eqnarray}
where $\phi=BS$, represents the flux trough the unit cell, $S= |y_2|b_1+2 
\frac{|y_2| b_2}{2}+ 2 \frac{y_1 b}{4}$.

Taking these results into account, the following hopping terms are present in 
the Hamiltonian:
\begin{eqnarray}
&& t_{32}^{\uparrow,\uparrow} = t_h e^{i\varphi_1} , \varphi_1=  \frac{2 \pi }{\phi_0} 
(-By_2)
\nonumber\\
&& t_{21}^{\uparrow,\uparrow} = t e^{i\varphi_2}, t_{43}^{\uparrow,\uparrow}= t 
e^{i\varphi_2},  t_{21}^{\uparrow,\downarrow} = \lambda e^{i\varphi_2}, 
t_{43}^{\uparrow,\downarrow}= \lambda e^{i\varphi_2} , \varphi_2 = \frac{2 \pi }{
\phi_0} B \frac{ |y_2| b_2}{2},
\nonumber\\
&& t_{54}^{\uparrow,\uparrow} = t e^{i\varphi_3}, t_{15}= t e^{i\varphi_3} , t_{54}^{
\uparrow,\downarrow} = -\lambda e^{i\varphi_3}, t_{15}^{\uparrow,\downarrow}= -\lambda 
e^{i\varphi_3} , \varphi_3 = \frac{2 \pi }{\phi_0} B \frac{ y_1 b}{4},
\label{Aequ4}
\end{eqnarray}
while $t_{i,j}^{\uparrow,\uparrow}=t_{i,j}^{\downarrow,\downarrow}$, 
$t_{i,j}^{\uparrow,\downarrow}=-t_{i,j}^{\downarrow,\uparrow}$ holds, 
since we have taken only
Rashba spin-orbit interactions into account \cite{E54}. 

\section{The secular equation}

The Secular equation Eq.(\ref{equ11}) in which Eq.(\ref{equ10}) has been 
introduced can be mathematically reduced to the diagonalization of the following
$4 \times 4$ matrix: 
\begin{equation}
	\begin{pmatrix}
	A_f & e^{-ikb}(t_c e^{ika}-W_{1}) & 0 &  e^{-ikb}( \lambda_c e^{ika}-W_{2} )\\
	
	e^{ikb}(t_c e^{- ika}-W_{1}^* ) & A_f & e^{ikb}(-\lambda_c e^{- ika} + W_{2}^* ) & 0\\
	
	0 & e^{-ikb}(-\lambda_c e^{  ika} + W_{2}) & A_f & e^{-ikb}(t_c e^{ ika}-W_{1} )\\
	
	e^{ikb}(\lambda_c e^{ - ika}-W_{2}^*) & 0 & e^{ikb}(t_c e^{ - ika}-W_{1}^*) & A_f
	\end{pmatrix} ,
\end{equation}
where $W_{1f}= (\lambda^2 -t^2) \left( \frac{1}{\bar{\epsilon}_2 
\bar{\bar{\epsilon}}_2} t_h e^{-i\varphi} - \frac{1}{\bar{\bar{\epsilon}}_3}
\right)$,  $W_{2f}= 2\lambda t \left( \frac{1}{\bar{\epsilon}_2 
\bar{\bar{\epsilon}}_2} t_h e^{-i\varphi} - \frac{1}{\bar{\bar{\epsilon}}_3}
\right)$.

Furthermore the $A$ and $V$ expressions present in Eq.(\ref{equ12})
are defined as:
\begin{eqnarray}
A = A_f &-& \frac{1}{A_f} \left( \lambda_c e^{i(\varphi_k - 2 \varphi_3)} + 2 
\lambda t 
\xi \right)  \left( \lambda_c e^{-i(\varphi_k - 2 \varphi_3)} + 2 \lambda t 
\xi^* \right) 
\nonumber\\
&-& \frac{1}{A_f}  \left( t_c e^{i(\varphi_k - 2 \varphi_3)} - (t^2 - \lambda^2) 
\xi \right) \left( t_c e^{-i(\varphi_k - 2 \varphi_3)} - (t^2 - \lambda^2) 
\xi^* \right),
\label{Bequ2}
\end{eqnarray}
where
\begin{eqnarray}
\xi = \left( \frac{\bar{\epsilon}_4}{\bar{\epsilon}_3 \bar{\epsilon}_4-t_f^2 } 
e^{i \varphi} - \frac{t_h}{\bar{\epsilon}_2^2 - t_h^2} \right)
\label{Bequ3}
\end{eqnarray}
and
\begin{eqnarray}
V = &-& \frac{1}{A_f} \left( \lambda_c e^{i(\varphi_k - 2 \varphi_3)} + 2 \lambda t 
\xi \right) \left( t_c e^{-i(\varphi_k - 2 \varphi_3)} - (t^2 - \lambda^2) 
\xi^* \right) -
\nonumber\\
&-& \frac{1}{A_f} \left( -\lambda_c e^{-i(\varphi_k - 2 \varphi_3)} - 2 \lambda t 
\xi^* \right) \left( t_c e^{i(\varphi_k - 2 \varphi_3)} - (t^2 - \lambda^2) \xi 
\right).
\label{Bequ4}
\end{eqnarray}

\section{The flat band conditions derived from the $(A -i V)$ 
expression}

The studied expression can be written as:
\begin{eqnarray}
(A - i V)= T_0+ \bar{T}_1  \cos(\varphi_k) + \bar{T}_2 
\sin(\varphi_k)=0,
\label{Cequ1}
\end{eqnarray}
where $T_0$ is the same as seen in (\ref{equ14}), and $\bar{T}_1, 
\bar{T}_2$ are given by
\begin{eqnarray}
&&\bar{T}_1= \frac{1}{A_f} \bigg(
-\cos(7 \varphi_3) \frac{2 (2 \lambda t \lambda_c  + t_c (\lambda^2 - t^2))}{ \bar{\bar{\epsilon}}_3}
+ \sin(7 \varphi_3) \frac{2 (-2 \lambda t t_c + \lambda_c (\lambda^2 - t^2))}{\bar{\bar{\epsilon}}_3} +
\nonumber\\
&&\hspace{0.5cm} + \cos(2 \varphi_3) \frac{2 (2 \lambda  t \lambda_c + t_c (\lambda^2 - t^2 )) t_h}{\bar{\epsilon}_2 \bar{\bar{\epsilon}}_2} 
- \sin(2 \varphi_3) \frac{2 (-2 \lambda t t_c + \lambda_c (\lambda^2 - t^2)) t_h}{\bar{\epsilon}_2 \bar{\bar{\epsilon}}_2}
\bigg) =0,
\nonumber\\
&&\bar{T}_2=\frac{1}{A_f}\bigg( 
-\cos(7 \varphi_3) \frac{2 (-2 \lambda t t_c + \lambda_c (\lambda^2 - t^2))}{ \bar{\bar{\epsilon}}_3}
-\sin(7 \varphi_3) \frac{2 (2 \lambda  t \lambda_c + t_c (\lambda^2 - t^2 ))}{ \bar{\bar{\epsilon}}_3} +
\nonumber\\
&&\hspace{0.5cm} + \cos(2 \varphi_3)  \frac{2 (-2 \lambda t t_c + \lambda_c (\lambda^2 - t^2)) t_h}{\bar{\epsilon}_2 \bar{\bar{\epsilon}}_2} 
+ \sin(2 \varphi_3) \frac{2 (2 \lambda  t \lambda_c + t_c (\lambda^2 - t^2)) t_h}{\bar{\epsilon}_2 \bar{\bar{\epsilon}}_2} \bigg) =0.
\label{Cequ2}
\end{eqnarray}
Using the same notation as in Eq.(\ref{equ20}) this expression provides:
\begin{eqnarray}
\bar{T}_1 = \frac{1}{A_f} (-K_g v + S_g u)=0, \quad \bar{T}_2 =  \frac{1}{A_f} 
(-K_g u - S_g v )=0.
\label{Cequ3}
\end{eqnarray}
As in the case of Eq.(\ref{equ20}), only the solution $K_g=S_g=0$ exists,
hence we reobtain the solutions derived from the $(A + i V)$ expression. This 
is an important result because of the following reason: it is known that  
usually, the many-body spin orbit interaction breaks the spin-projection 
double degeneracy of each band \cite{E38,E41}.
But here one observes, that if one creates a 
flat band using SOI, the flat band will remain double degenerated.  

\pagebreak

\section{Sets of Hamiltonian parameter data.}

\begin{table}[htb!]
	\begin{tabular}{|c|c|c|c|c|c||c|c|l|c|c|c|c|c|c||c|c|}
		\cline{1-8} \cline{10-17}
		\multicolumn{8}{|c|}{Set.1} &  & \multicolumn{8}{c|}{Set.2} \\ \cline{1-8} \cline{10-17} 
		\multicolumn{6}{|c||}{\begin{tabular}[c]{@{}c@{}}The unrestricted\\  parameters\end{tabular}} & \multicolumn{2}{c|}{\begin{tabular}[c]{@{}c@{}} The parameters   \\calculated from the \\  flat band conditions\end{tabular}} &  & \multicolumn{6}{c||}{\begin{tabular}[c]{@{}c@{}}The unrestricted\\  parameters\end{tabular}} & \multicolumn{2}{c|}{\begin{tabular}[c]{@{}c@{}} The parameters   \\calculated from the \\  flat band conditions\end{tabular}} \\ \cline{1-8} \cline{10-17} 
		$\epsilon_1$     & $\epsilon_2$     & $\epsilon_3$     & $\epsilon_4$    & $t$    & $t_h$    & $t_c^{rfbc}$  & $t_f^{rfbc}$  &  & $\epsilon_1$     & $\epsilon_2$     & $\epsilon_3$    & $\epsilon_4$    & $t$    & $t_h$    & $ t_c^{rfbc}$  & $  t_f^{rfbc}$  \\ \cline{1-8} \cline{10-17} 
		0.17   & 0.49  & 0.22  & 3.36  & \hspace{0.05cm} 1 \hspace{0.05cm}      & 1.5      &  \hspace{0.3cm} 1.16  \hspace{0.3cm}  & 2.29  &  & 0.87  & 1.22          & 0.82            & 0.36            & \hspace{0.05cm} 1 \hspace{0.05cm}      & 0.9      &  \hspace{0.3cm} 2.21  \hspace{0.28cm} & 0.14   \\ \cline{1-8} \cline{10-17} 
	\end{tabular}
\caption{The Set.1 and Set.2 of Hamiltonian parameter data.}
\end{table}	


\begin{table}[htb!]
	\begin{tabular}{|c|c|c|c|c|c||c|c|l|c|c|c|c|c|c||c|c|}
		\cline{1-8} \cline{10-17}
		\multicolumn{8}{|c|}{Set.3}                                                                                                                                                                        &  & \multicolumn{8}{c|}{Set.4}                                                                                                                                                                        \\ \cline{1-8} \cline{10-17} 
		\multicolumn{6}{|c||}{\begin{tabular}[c]{@{}c@{}}The unrestricted\\  parameters\end{tabular}} & \multicolumn{2}{c|}{\begin{tabular}[c]{@{}c@{}}The parameters   \\calculated from the \\  flat band conditions\end{tabular}} &  & \multicolumn{6}{c||}{\begin{tabular}[c]{@{}c@{}}The unrestricted\\  parameters\end{tabular}} & \multicolumn{2}{c|}{\begin{tabular}[c]{@{}c@{}}The parameters   \\calculated from the \\  flat band conditions\end{tabular}} \\ \cline{1-8} \cline{10-17} 
		$\epsilon_1$     & $\epsilon_2$     & $\epsilon_3$     & $\epsilon_4$    & $t$    & $t_h$    & $t_c^{rfbc}$                                       & $t_f^{rfbc}$                                      &  & $\epsilon_1$     & $\epsilon_2$     & $\epsilon_3$    & $\epsilon_4$    & $t$    & $t_h$    & $t_c^{rfbc}$   & $t_f^{rfbc}$ \\ \cline{1-8} \cline{10-17} 
		0.11   & 0.10   & 0.92  & 0.86  & \hspace{0.05cm} 1 \hspace{0.05cm} & 0.85     &\hspace{0.31cm} 1.44 \hspace{0.31cm} & 1.23   &  & 0.65  & 0.49  & 1.4  & 0.86  & \hspace{0.05cm} 1 \hspace{0.05cm}  & \hspace{0.05cm} 2   \hspace{0.05cm}      & \hspace{0.31cm}1.31 \hspace{0.31cm}                                           & 1.68                                           \\ \cline{1-8} \cline{10-17} 
	\end{tabular}
\caption{The Set.3 and Set.4 of Hamiltonian parameter data.}
\end{table}

We note that in the Sets 1-4 all parameters presented are given in $t$ units, 
and the rigidly fixed flat band conditions have been deduced at 
$B=\lambda=\lambda_c=0$. In the cases of the Sets 2 and 4,
when $B \ne 0$ plots are done, the $B$ value was
deduced from Eq.(22), i.e. $I_{\varphi}=X_{\varphi}=-1$. E.g., for $\varphi_b =
3 \varphi_3$ (regular pentagon), one has at minimum $B$ the relation 
$\varphi_3= (1/5) \pi$. For connection to $B$, see also
Eq.(A4). 

\section{Dispersive bands from flat bands}

Let us consider that at
$B=\lambda=\lambda_c=0$ one modifies the $t_c=t_c^{rfbc}$
rigid flat band condition value. What happens to the band
is exemplified in Fig.13, where $8.62\%$ modification has been taken into 
account relative to $t_c^{rfbc}$. As seen, the flat band becomes a dispersive
band.
\begin{figure}[!htb]
        \includegraphics[width=0.5\textwidth]{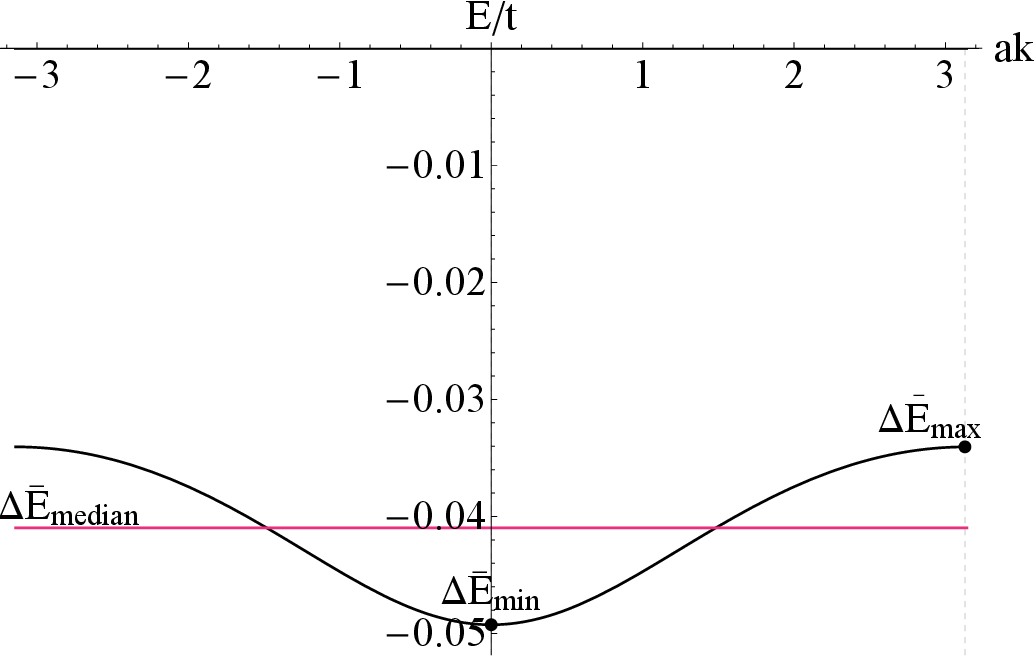}
	\caption{The flat band originally placed at the origin of the
energy axis, becomes dispersive under the action of $\Delta t_c$ at zero SOI
couplings. The line in the middle of the dispersive band (at position 
$\Delta \bar E_{median}$) shows the median of the band, while 
$\Delta \bar{E}_{min}$ and $\Delta \bar{E}_{max}$ denote the minimum and maximum
position in the dispersive band relative to the median. The 
$\Delta \bar{E}_{\alpha}$, $\alpha=median, min, max$ values 
are indicated in $t$ units. For exemplification we have used the Set 1. of 
Hamiltonian parameter data from Appendix D.}
	\label{fig13}
\end{figure}
How $\Delta \bar E_{median}$, $\Delta \bar{E}_{min}$ and $\Delta \bar{E}_{max}$
change as function of $\Delta t_c$ is exemplified in Fig.14.
\begin{figure}[!htb]
        \includegraphics[width=\textwidth]{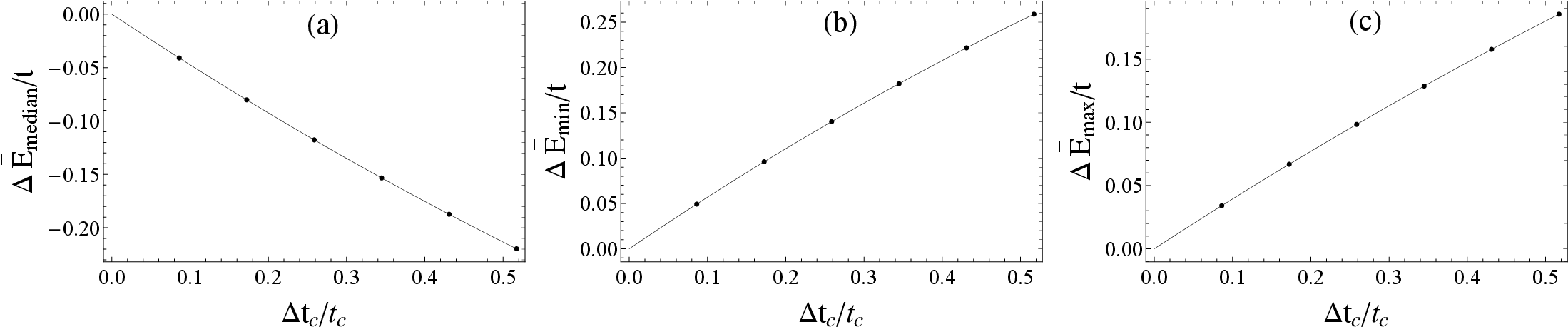}  
	\caption{The different $\Delta \bar E_{\alpha}$ quantities from Fig.12 
as function of $\Delta t_c$ normalized to $t_c=t_c^{rfbc}$. The presented 
cases: (a) the median- ($\Delta \bar{E}_{median}$),  (b) the maximum- 
($\Delta \bar{E}_{max}$), and (c) the minimum ($\Delta \bar{E}_{min}$).
For exemplification we have used the Set 1. of Hamiltonian parameter data from 
Appendix D.}
	\label{fig14}
\end{figure}
One notes that the original flat band placed in the origin of the energy axis
was double degenerated relative to the spin projection, and since
$\lambda=\lambda_c=0$, this double 
degeneracy remains valid also in the case of the dispersive band emerging at
$\Delta t_c \ne 0$. 

\end{document}